\documentclass[a4paper,11pt]{article}
\pdfoutput=1 

\usepackage{jheppub} 

\usepackage{mathrsfs,amsmath,amsthm,latexsym,amssymb,amsfonts,epsfig,cancel,enumerate,graphicx,subcaption,txfonts,color,enumerate}
\usepackage{subcaption}
\usepackage{amsmath}
\usepackage{graphicx}
\usepackage{dcolumn}
\usepackage{bm}
\usepackage{hyperref}

\usepackage{tikz}
\usepackage{bm}
\newcommand*{\circled}[1]{\lower.7ex\hbox{\tikz\draw (0pt, 0pt)%
    circle (.5em) node {\makebox[1em][c]{\small #1}};}}

\usepackage{subcaption}

\title{Shadows and Observational Images of a Schwarzschild-like Black Hole Surrounded by a Dehnen-type Dark Matter Halo}

\author[1]{Zuting Luo}
\author[1]{Meirong Tang}
\author[1,*]{Zhaoyi Xu}

\affiliation[1]{College of Physics, Guizhou University, Guiyang 550025, China}

\emailAdd{zyxu@gzu.edu.cn(Corresponding author)}

\abstract{
This paper investigates the optical appearance of a Schwarzschild-like black hole (BH) surrounded by a Dehnen-(1, 4, 5/2)  type dark matter (DM) halo, with a focus on how the DM halo's density $\rho_{s}$ and radius $r_{s}$ influence the BH's shadow and photon ring. First, the radius $r_h$ of the BH's event horizon and the equation of motion for photons were derived, and observational data from the Event Horizon Telescope (EHT) for M87* were used to constrain the parameters $\rho_{s}$ and $r_{s}$ of the DM halo. Afterward, by varying the values of $\rho_{s}$ and $r_{s}$, key parameters such as the effective potential $V_{eff}$ of photons, the critical impact parameter $b_{ph}$, the radius $r_{isco}$ of the innermost stable circular orbit, and the radius $r_{ph}$ of the photon sphere were calculated for each case. It was found that as $\rho_{s}$ and $r_{s}$ increase, the above mentioned parameters all show an increasing trend. Subsequently, we investigated the optical appearance of the BH illuminated by two types of accretion models: optically and geometrically thin disk models and spherical accretion models. The findings indicate that as $\rho_{s}$ and $r_{s}$ increase, the peak of the received intensity shifts toward a higher impact parameter $b$, resulting in a distinct optical appearance.
}


\begin{document}
\maketitle
\flushbottom

\section{\label{sec:level1}Introduction}
General relativity (GR) stands as one of the most successful theories in describing gravitational phenomena and has gained widespread acceptance. As one of its key theoretical predictions, BHs have remained a focal point of physics research for decades, captivating scientific attention due to their enigmatic nature. Extensive theoretical investigations have been conducted, as exemplified by studies in references \cite{LIGOScientific:2016aoc,Berti:2009kk,LIGOScientific:2016lio,EventHorizonTelescope:2019dse,Banados:1992wn,Kovtun:2004de,Hawking:1976de,Hawking:1975vcx,EventHorizonTelescope:2022wkp,Shenker:2013pqa} which explore BHs from diverse perspectives. As one of the possible products of the gravitational collapse of massive stars \cite{Oppenheimer:1939ue}, BHs cannot be directly observed due to their extremely strong gravitational fields, and thus can only be studied through indirect approaches.
The study of BHs entered a new phase in 2019 when the EHT team unveiled the image of the supermassive BH located at the center of a galaxy \cite{EventHorizonTelescope:2019uob, EventHorizonTelescope:2019ggy}. The image of BH M87* revealed a central dark area encircled by a luminous ring. The shaded region in the figure corresponds to the BH shadow, which appears as a dark area in the observer's field of view after light rays fall into the BH horizon. The bright ring structure in the image includes the BH photon ring. The photon ring is formed because the gravitational field of the BH itself causes the distortion of the spacetime structure around it, causing the trajectories of photons passing near the BH to bend, thereby producing a bright ring structure in the observer’s field of view \cite{Gralla:2019xty}. This gravitational lensing effect, where light rays are deflected around the BH, not only demonstrates the profound influence of the BH on the surrounding spacetime but also provides a way to further study the properties of BHs. 
The release of real BH images has not only inspired theoretical research on BH imaging but also led to extensive subsequent studies. Subsequently, comprehensive investigations of BH images under various gravitational effects have been conducted \cite{Amir:2016cen,Hioki:2009na,Abdujabbarov:2016hnw,Chakhchi:2022fls,Zeng:2022fdm,Zeng:2022pvb,Yang:2023tip,Promsiri:2023rez,Meng:2023uws,Zeng:2023zlf,EslamPanah:2024gxx,Gallo:2024wju,Kala:2024fvg,Meng:2024puu,Nojiri:2024qgx,Zeng:2024ptv,Wang:2025ihg,Zeng:2025kqw,Yashwanth:2024suw}.

Although GR has achieved great success in many aspects of describing gravity, it still does not include DM and dark energy. Observations of the cosmic microwave background indicate that the universe is mainly composed of $26.8\%$ DM and $68.5\%$ dark energy \cite{Planck:2018vyg,WMAP:2012nax}. Therefore, it is natural to assume that BHs do not exist completely in isolation, and there will also be DM surrounding them. Although DM has not been directly detected yet, there are already a lot of astronomical observation evidences indirectly proving its existence, such as the rotation curves of galaxy disks \cite{Rubin:1980zd,Lelli:2016zqa}, the Bullet Cluster \cite{Clowe:2006eq,Corbelli:1999af}, baryon acoustic oscillations \cite{SDSS:2005xqv,WMAP:2010qai}, and so on. Therefore, the combination of DM and BHs has already become an object of great concern and high research value at present. In the study of interactions between DM and BHs, various analytical models have been proposed to describe DM BH solutions \cite{Graham:2005xx,Dutton:2014xda,Navarro:1995iw,Burkert:1995yz,Dehnen:1993uh,Gohain:2024eer,Pantig:2022whj}. One widely used model is the Dehnen model \cite{Dehnen:1993uh}. Proposed by Dehnen in 1993, this model features a spherically symmetric density profile determined by three parameters ($\alpha$, $\beta$, $\gamma$), designed to simulate the structure and dynamics of elliptical galaxies and galactic bulges \cite{Dehnen:1993uh}. The gravitational potential corresponding to the Dehnen model is analytical for all values of $\gamma$ . In the specific case where $\gamma = 3/2$, the Dehnen model most closely matches de Vaucouleurs' $R^{1/4}$ empirical profile—used to describe the surface brightness of elliptical galaxies and bulges—in terms of surface density and distribution function \cite{Dehnen:1993uh}. Consequently, the Dehnen model has been widely extended to construct DM halos surrounding BHs. Extensive studies have been conducted on the properties of BH solutions surrounded by Dehnen-type DM halos, including thermodynamic characteristics and quasinormal modes \cite{Alloqulov:2025ucf,Al-Badawi:2025njy,Al-Badawi:2024qpt,Jha:2024ltc}. Moreover, DM does not participate in electromagnetic interactions in most scenarios  \cite{Konoplya:2019sns}.  As a result, the gravitational influence of DM on BHs cannot be directly probed through electromagnetic observations. Therefore, one studies the optical appearance of BHs in a DM halo to explore DM's influence on BHs. Relevant research has been conducted for this purpose \cite{Jusufi:2019nrn,Jha:2024ltc,Macedo:2024qky,Faraji:2024ein,Zeng:2021mok,Atamurotov:2021hoq}. 

The spacetime around a BH can be influenced by the surrounding matter field, and this influence will be reflected in the image of the BH \cite{Xu:2018wow,Haroon:2018ryd,Guo:2022rql,Jha:2024ltc,Macedo:2024qky}. In addition, since BHs accrete the surrounding luminous matter, the different distributions of these accreting materials will further affect the image of the BH. However, it is difficult to simulate a realistic accretion disk theoretically. Therefore, the accretion models used in the theoretical study of BH accretion imaging are all simplified. At present, several widely adopted accretion disk models include the optically thin static accretion disk model \cite{Wang:2023vcv, Sui:2023tje, Li:2021riw, Uniyal:2022vdu, Hu:2023bzy, Fathi:2023ccx, Yang:2024utv} and the spherical accretion model, which features a spherically symmetric distribution of the accreting matter. \cite{Li:2024abk,Abdujabbarov:2016hnw,Chakhchi:2022fls,Malik:2024foi,Hu:2022lek,Chen:2025ifv}. The optical appearance of BHs formed by these different accretion models can, to a certain extent, reflect the spacetime geometry around the BH, providing a new window for the verification of different gravitational theories. In recent years, the imaging of BHs under various accretion models has attracted significant research attention, as exemplified by the simulation studies of shadow images under different accretion scenarios in \cite{Hu:2022lek,Wen:2022hkv,Gao:2023mjb,Zeng:2021dlj,He:2021htq,Huang:2024bbs,Zare:2024dtf,Kumaran:2023brp,Meng:2023htc,DeMartino:2023ovj,Meng:2024puu,Aslam:2024mhn,Wang:2023rjl,Zeng:2023fqy,Uniyal:2023ahv,Li:2024owp,Wang:2024lte,Fathi:2023ccx}.

Recently, Ahmad Al-Badawi et al. proposed a novel Schwarzschild-like solution describing an asymptotically flat Schwarzschild BH surrounded by Dehnen-(1, 4, 5/2) type DM \cite{Al-Badawi:2024asn}. After analyzing the curvature properties and energy conditions of this solution to investigate the BH characteristics, they conducted a systematic study of the timelike geodesics around the BH, laying the foundation for subsequent research on the imaging features of accretion disks around such BHs. Follow-up studies have systematically examined other properties of this BH \cite{Liang:2025vux,Hamil:2025pte,Rani:2025esb}. Among them, Rani et al. \cite{Rani:2025esb} employed the ray-tracing method to investigate the optical appearance of the BH illuminated by an optically thin accretion disk. However, their work only considered a specific thin-disk model (with a radiation profile following a Gaussian function) and did not incorporate existing observational data to constrain the DM parameters. This motivates us to conduct a more comprehensive study on the optical appearance of this BH solution under different static accretion disk and spherical accretion models, while constraining the DM parameters with available observational data. By providing optical images under various accretion scenarios, we aim to extend the understanding of the observational signatures of this BH model.

Therefore, the purpose of this paper is to investigate the optical appearance of a Schwarzschild-like BH immersed in Dehnen-type DM under both static thin disk accretion models and spherical accretion scenarios, and to explore the influence of DM halos on BHs through these optical signatures. Specifically, we first examine the photon sphere around a Schwarzschild-like BH solution surrounded by a Dehnen-type DM halo \cite{Al-Badawi:2024asn}. Using the ray-tracing method, we investigate the photon trajectories around the BH and constrain the DM parameters with existing EHT observational data of M87*. Within these constrained parameter ranges, we classify the photon trajectories around this Schwarzschild-like BH. Finally, we analyze the optical appearance under both optically and geometrically thin accretion disks and spherical accretion scenarios.

The structure of this paper is as follows: In Section \ref{sec:level2} , we recall the Schwarzschild-like BH surrounded by a Dehnen-type DM halo. In Section \ref{sec:level3} , we calculate the equations of motion of photons along geodesics and constrain the parameters $\rho_s$ and $r_s$ of the DM halo using the observational data of M87*. In Section \ref{sec:level4} , we classify the geodesic trajectories following the method in \cite{Gralla:2019xty}, present the corresponding photon trajectory diagrams for different values of parameters $\rho_{s}$ and $r_{s}$, and study the Schwarzschild-like optical appearance of a BH surrounded by a Dehnen-type DM halo under an optically and geometrically thin accretion disk. In Section \ref{sec:level5}, we study the optical images of a BH under thin spherical accretion models. Finally, in Section \ref{sec:level6} , we present the conclusions. In this work, we employ natural units where the speed of light $c$ and the gravitational constant 
$G$ are both set to unity $(c = G = 1)$. Additionally, the BH mass $M$ is set to 1 for computational purposes. The metric convention \((-, +, +, +)\) is adopted throughout the paper.

\section{\label{sec:level2}A brief recall of Schwarzschild-like black hole}
\begin{figure*}[htb]
	\includegraphics[width=1\textwidth]{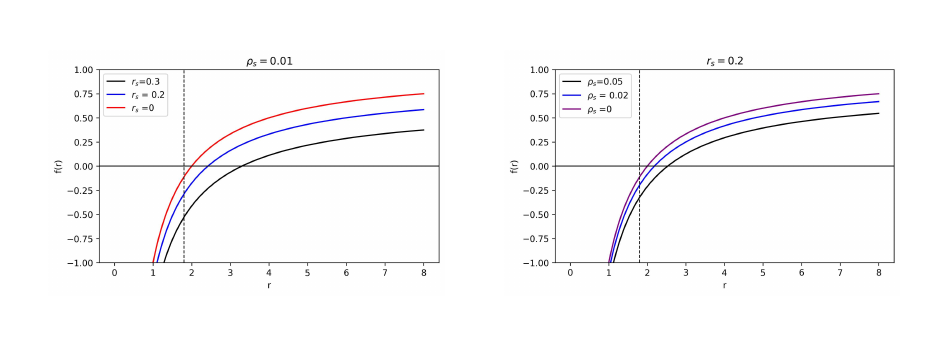}
	\caption{The metric function $f(r)$ versus $r$ at fixed $\rho_{s}=0.01$ (left panel) and versus $\rho_{s}$ at fixed $r_{s}=0.2$ (right panel).} 
	\label{function}
\end{figure*}

The metric of this Schwarzschild-like BH surrounded by a Dehnen-(1, 4, 5/2) type  DM halo can be described by the following line element:
\begin{equation}\label{1}
	\begin{split}
		ds^{2} = -f(r)dt^{2} + \frac{1}{f(r)}dr^{2} + r^2\left(d\theta^{2} + \sin^{2}\theta \, d\phi^{2}\right)
	\end{split},
\end{equation}
where \cite{Al-Badawi:2024asn}
\begin{equation}\label{2}
	\begin{split}
		f(r) &= 1 - \frac{2M}{r} - 32\pi\rho_s r_s^2 \sqrt{\frac{r + r_s}{r}}
	\end{split}.
\end{equation}
Among them, $M$ is the mass of the BH, $\rho_{s}$ and $r_{s}$ are the density and radius of the central DM halo respectively. It can be seen from Eq.~\eqref{2} that $f(r)$ is a function of $\rho_{s}$ and $r_{s}$. We have drawn the function graph of $f(r)$ when $\rho_{s}$ and $r_{s}$ take certain values in Fig. \ref{function}. As clearly shown in Fig. \ref{function} , when fixing $\rho_{s}$ and $r_{s}$ separately, the $f(r)$ versus $r$ curve shifts progressively rightward as these parameters increase.

\begin{figure*}[htb]
	\includegraphics[width=1\textwidth]{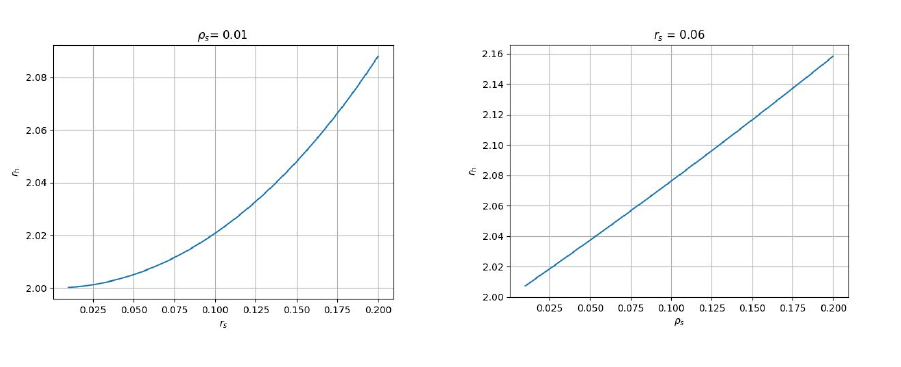}
	\caption{The variation of the event horizon radius with $\rho_{s}$ and $r_{s}$: left panel shows fixed $\rho_{s} = 0.01$; right panel shows fixed $r_{s} = 0.06$. } 
	\label{rh}
\end{figure*}

The event horizon radius $r_{h }$ is determined as the solution to $f(r)=0$. We derive the specific expression for $r_h$ as follows:
\begin{equation}\label{3}
	\begin{split}
		r_h = \frac{2M + 512\pi^2 \rho_s^2 r_s^5+32\pi\sqrt{2M\rho_s^2 r_s^4 (2M + r_s) + 256\pi^2 \rho_s^4 r_s^{10}}}{1 - 1024\pi^2 \rho_s^2 r_s^4} 
	\end{split}.
\end{equation}
Then, Fig. \ref{rh} displays the dependence of the event horizon radius on both $\rho_{s}$ and $r_{s}$ parameters.  The metric functions and the variation of the horizon radius with the DM parameters are shown in Figures 1 and 2, respectively, which are consistent with the results reported in \cite{Al-Badawi:2024asn}. Therefore, we will not discuss them in further detail here.

\section{\label{sec:level3}Photon spheres of Schwarzschild-like black hole}
\subsection{\label{sec:level3.1}The trajectory of photon motion and the effective potential}
To investigate photon motion near a Schwarzschild-like BH surrounded by a Dehnen-type DM halo, we must examine the geodesic equations of photons near the BH. Starting from the Lagrangian of photons, we have
\begin{equation}\label{4}
	\begin{split}
		\mathcal{L} = \frac{1}{2} g_{ij} \frac{dx^i}{d\tau} \frac{dx^j}{d\tau}
	\end{split},
\end{equation}
where $\tau$ represents the affine parameter. For the line element  \eqref{1}, we obtain 
	\begin{equation}\label{5}
		\begin{split}
			\mathcal{L} = \frac{1}{2} \left[ 
			-f(r) \dot{t}^2 
			+ \frac{\dot{r}^2}{f(r)} 
			+ r^2 \dot{\theta}^2 
			+ \left(r^2 \sin^2 \theta\right) \dot{\phi}^2 
			\right].
		\end{split}
	\end{equation}
Here, the dots denote derivatives with respect to $\tau$. In a static spherically symmetric spacetime, two conserved quantities arise due to symmetries, namely the energy $E$ and the angular momentum $L$, which can be expressed as follows
\begin{equation}\label{6}
	\begin{split}
		E \equiv - \frac{\partial \mathcal{L}}{\partial \dot{t}} = f(r) \dot{t}
	\end{split},
\end{equation}
\begin{equation}\label{7}
	\begin{split}
		L \equiv \frac{\partial \mathcal{L}}{\partial \dot{\phi}} = r^2 \sin^2 \theta \; \dot{\phi}
	\end{split}. 
\end{equation}

\begin{figure*}[htb]
	\includegraphics[width=1.1\textwidth]{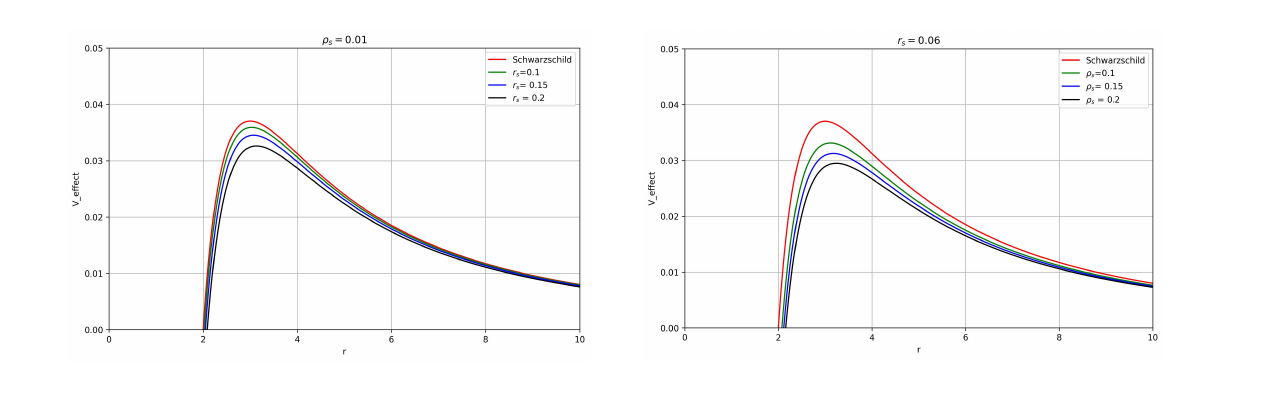}
	\caption{Effective potential curves for different values of $\rho_{s}$ and $r_{s}$. Left: fixed $\rho_{s} = 0.01$; right: fixed $r_{s} = 0.06$. Red, green, blue, and black lines correspond to Schwarzschild, and $r_{s}$ or $\rho_{s} = 0.1$, $0.15$, $0.2$, respectively.}
	\label{veff}
\end{figure*}

For photon,  the Lagrangian $\mathcal{L} = 0$. Considering the symmetry of the spacetime, we restrict the motion of the photon to the equatorial plane ($\theta = \pi/2$) for convenience. After redefining the affine parameter as $\tau = \tau / L$, we obtain the equations of motion for photons
\begin{equation}\label{tdot}
	\begin{split}
		\dot{t} = \frac{1}{b\,f(r)}
	\end{split},
\end{equation}
\begin{equation}\label{phidot}
	\begin{split}
		\dot{\phi} = \pm \frac{1}{r^2}
	\end{split}, 
\end{equation}
\begin{equation}\label{rdot}
	\begin{split}
		\dot{r}^2 = \frac{1}{b^2} - V_{eff}(r)
	\end{split}. 
\end{equation}
Here, the symbol ``$\pm$'' denotes the propagation direction of photons. The ``$+$'' corresponds to the photon moving in the clockwise direction, while the ``$-$'' represents the photon moving in the counterclockwise direction. Moreover, $b \equiv L/E$ denotes the impact parameter of the photon, and $V_{eff}$  refers to the effective potential that governs the radial motion of the photon. This can be mathematically expressed as follows:
\begin{equation}\label{11}
	\begin{split}
		V_{eff}(r) = \frac{f(r)}{r^2}
	\end{split}. 
\end{equation}
By combining Eqs.~\eqref{phidot} and ~\eqref{rdot}, we obtain the trajectory equation of the photon
\begin{equation}\label{rphi}
	\begin{split}
		\left( \frac{dr}{d\phi} \right)^2 = r^4\left[\frac{1}{b^2} -V_{eff}(r)\right]
	\end{split}. 
\end{equation}
It can be seen from this that the trajectory of photons exhibits a close correlation with the impact parameter $b$ and the effective potential $V_{eff}$. Photons with different impact parameters near a BH exhibit three types of trajectories: scattering to infinity, falling into the BH, and a critical trajectory in between--the one that forms an unstable photon sphere. For photons on unstable orbits, any perturbation will cause them to either plunge into the BH or escape to infinity, which corresponds to the boundary of the BH shadow as observed by a distant observer. For photons located on the unstable photon sphere, their effective potential and impact parameter must satisfy:
\begin{equation}\label{13}
	\begin{split}
		\left.V_{{eff}}(r)\right|_{r = r_{ph}}=\frac{1}{b_{ph}^2},\left.V_{{eff}}'(r)\right|_{r = r_{ph}} = 0
	\end{split}. 
\end{equation}
Here, $r_{ph}$ and $b_{ph}$ represent the radius of the unstable photon sphere and the critical impact parameter respectively. 
And the prime notation denotes derivative with respect to $r$. Hence, we obtain
\begin{equation}
	\begin{split}
		b_{ph}=\frac{r_{ ph}}{\sqrt{f(r_{ ph})}}\label{bph}.
	\end{split}
\end{equation}
In Fig. \ref{veff}, the corresponding curves of the effective potential varying with different values of $\rho_{s}$ and $r_{s}$ are presented.
It is evident from Fig.~\ref{veff}\; that the extrema of the effective potential curve correlate with the critical impact parameter $b_{ph}$, and the corresponding value of $r$ at the extremum represents the photon sphere radius $r_{ph}$. Moreover, when the value of $\rho_s$ (or $r_s$) is fixed, both $r_{ph}$ and $b_{ph}$ exhibit a gradually increasing trend as the other variable increases. 

\subsection{\label{sec:level3.2}Constraints of the M87* shadow on parameters}
In the previous subsection, we discussed the photon sphere radius $r_{ph}$ and the critical impact parameter $b_{ph}$ of the BH, both of which vary with $\rho_s$ and $r_s$. The variation of $b_{ph}$ reflects changes in the apparent size of the BH in the observer's frame, thereby offering a valuable opportunity to constrain the possible parameter ranges through observational data of BH shadows. The shadow radius $r_{sh}$ of the BH, as measured by a distant observer located at infinity, is given by
\begin{equation}\label{14}
	\begin{split}
		r_{sh} \equiv b_{ph}=\frac{r_{ph}}{\sqrt{f(r_{ph})}}
	\end{split}. 
\end{equation}

\begin{figure}[h]
	\centering
	\includegraphics[width=0.6\textwidth]{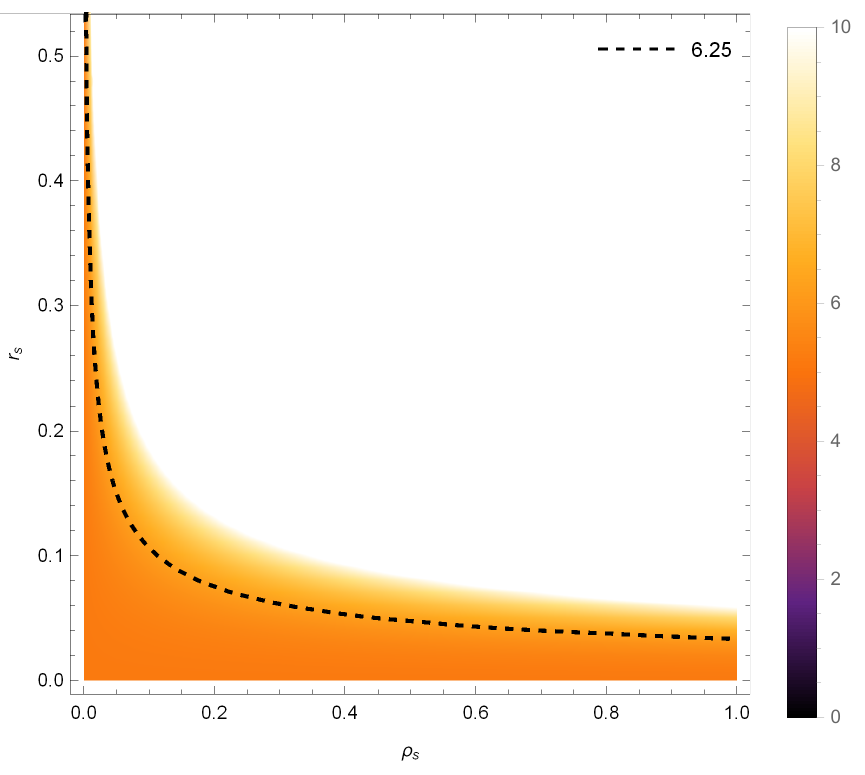}
	\caption{The constraints imposed by the M87* shadow radius on the DM halo density $\rho_{s}$ and radius $r_{s}$. The dashed line indicates the upper limit of the shadow radius (6.25), with the parameter space below this line representing the allowed region for $\rho_{s}$ and $r_{s}$. } 
	\label{fig:yueshu}
\end{figure}

In 2019, the EHT released the shadow image of the supermassive BH at the center of the M87* galaxy. The results showed that the angular diameter of the shadow is $\theta_{\rm M87^*} = (42 \pm 3)\ \mu\text{as}$ \cite{EventHorizonTelescope:2019ggy}.
The mass of M87* is $M_{\rm M87^*} = (6.5 \pm 0.9) \times 10^9 M_{\odot}$, where $M_{\odot}$ is the mass of the sun, and the distance from M87* to the Earth is $ D_{\rm M87^*} = 16.8\pm 0.8\, \text{Mpc} $. Moreover, the shadow diameter of M87* is calculated to be approximately $d_{sh}^{\rm M87^*}\simeq11.0 \pm 1.5$ \cite{Bambi:2019tjh}. This implies that the shadow radius $r_{sh}$ should satisfy: 
\begin{equation}\label{15}
	\begin{split}
		4.75\leq r_{sh}\leq 6.25
	\end{split}. 
\end{equation}
Therefore, in Fig.~\ref {fig:yueshu}, we present the constrained parameter space of $\rho_{s}$ and $r_{s}$ for a Schwarzschild-like BH surrounded by a Dehnen-type DM halo, based on the limits of the shadow radius of M87*. The black dashed line in Fig.~\ref {fig:yueshu}\; represents the BH shadow radius $r_{sh} = 6.25$. It should be noted that all values in the density plot exceed the Schwarzschild BH's shadow radius (i.e., $r_{sh} > 3\sqrt{3}$). Moreover, the region enclosed by this line and the $\rho_{s}$ and $r_{s}$ axes corresponds to shadow radii smaller than 6.25, indicating the allowed range of $\rho_{s}$ and $r_{s}$. Hence, the parameter values of $\rho_{s}$ and $r_{s}$ in the following analysis all lie within this range.

\section{\label{sec:level4}The images of a Schwarzschild-like black hole with static thin accretion disks}
In this section, we aim to study the optical appearance of a Schwarzschild-like BH surrounded by a Dehnen-type DM halo and illuminated by a static geometrically thin accretion disk. We assume that the observer is located at the North Pole, directly facing the accretion disk on the equatorial plane. Since photons emitted from different regions of the disk contribute differently to the BH image, we first classify the photon trajectories around the Schwarzschild-like BH and then analyze the resulting image.  The trajectories of photons have been partially investigated in \cite{Rani:2025esb}. Therefore, within the parameter ranges constrained by observational data, we focus here on briefly analyzing the influence of DM parameters on photon trajectories. Additionally, we employ three distinct thin disk accretion models to study the corresponding image features under their illumination.

\subsection{\label{sec:level4.1}Classification of photon trajectories: direct emission, lensing ring, photon ring}
Photons emitted from the accretion disk that reach a distant observer can equivalently be regarded as light rays traced backward from the observer that intersect the accretion disk near the BH. Each time a photon intersects with the accretion disk, it will extract energy from the accretion disk. Based on this, we adopt the classification method proposed in the literature \cite{Gralla:2019xty} and divide the photon trajectories into three categories based on the number of intersections \(n\) with the accretion disk. 

Here, we make the following assumption: $u=1/r$. After performing a transformation on Eq.~\eqref {rphi}, the photon trajectory equation can be reformulated in the following form:
\begin{equation}\label{16}
	\begin{split}
		\left(\frac{du}{d\phi}\right)^2 = \frac{1}{b^2} - f\left(\frac{1}{u}\right) u^2
	\end{split}. 
\end{equation}
Meanwhile, the The intersection count $n$ between the photon trajectory and the accretion disk can be characterized as follows:
\begin{equation}\label{17}
	\begin{split}
		n = \frac{\phi}{2\pi}
	\end{split},
\end{equation}
here $\phi$ denotes the total change in the azimuthal angle over the course of the entire photon trajectory. 
From Eq.~\eqref {16}, it can be known that the number of intersections $n$ is actually a function of the impact parameter $b$, and there is a specific relationship that they satisfy \cite{Peng:2020wun}:
\begin{equation}\label{18}
	\begin{split}
		n(b)=\frac{2m - 1}{4}, \quad m = 1,2,3,\cdots.
	\end{split}
\end{equation}
For a given value of the $m$, there exist $b_{m}^{\pm}$. Here $b_{m}^{-}<b_{ph}$ and $b_{m}^{+}>b_{ph}$. Hence,  null geodesics can be classified into three categories: direct emission, the lensing ring, and the photon ring. And they can be expressed as follows:
\begin{itemize}
	\item Direct emission: $1/4<n<3/4$ $\Leftrightarrow$ $b\in(0,b_2^-)\cup(b_2^+,\infty)$;
	\item Lensing ring: $3/4<n<5/4$ $\Leftrightarrow$ $b\in(b_2^-,b_3^-)\cup(b_3^+,b_2^+)$;
	\item Photon ring: $n>5/4$ $\Leftrightarrow$ $b\in(b_3^-,b_3^+)$.
\end{itemize}
Direct emission indicates that the photon intersects the accretion disk once, the lensing ring indicates that the photon intersects the accretion disk twice, and the photon ring indicates that the photon intersects the accretion disk three times or more. 
\begin{figure*}[htpb]
	\centering
	\includegraphics[width=0.46\textwidth]{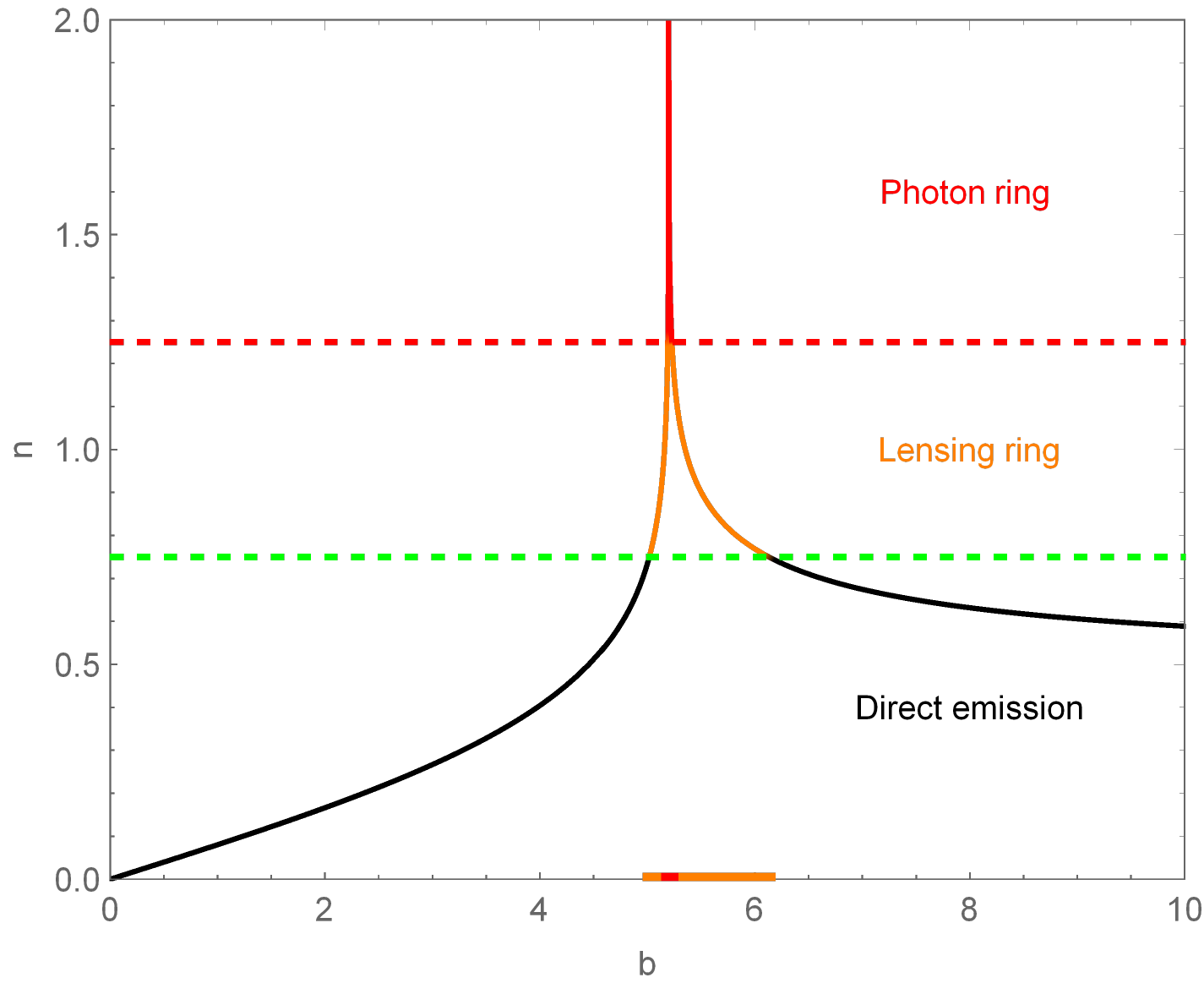}
	\includegraphics[width=0.42\textwidth]{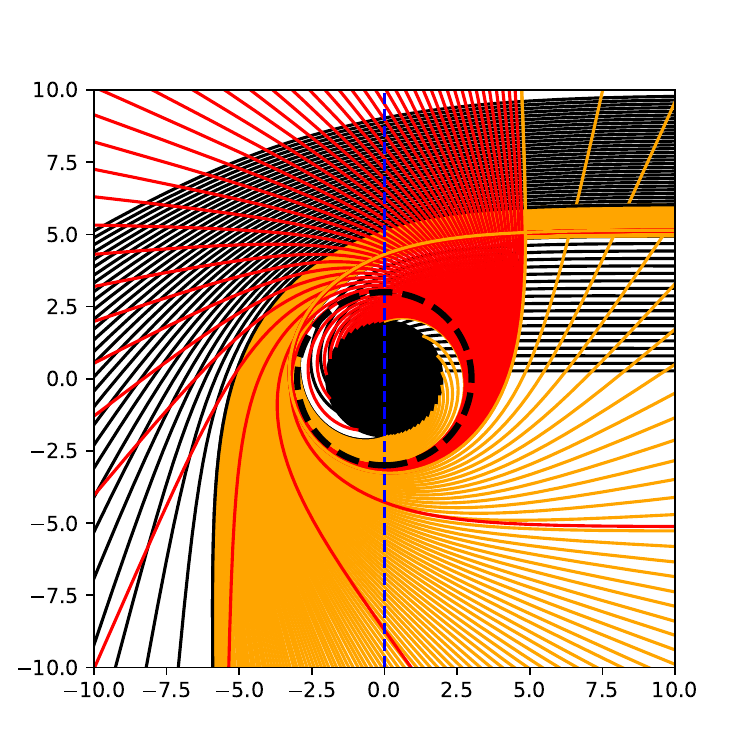}
	\caption{The intersection number curve for the Schwarzschild BH (left panel) and the corresponding photon trajectories (right panel). The black, orange, and red solid lines represent the direct emission, lensing ring, and photon ring, respectively. In the right panel, the blue dashed line denotes the equatorial plane of the BH, while the black disk and the black dashed circle indicate the BH and the photon sphere, respectively.}
	\label{sch}	
\end{figure*}
In Fig.~\ref{sch}, we take the Schwarzschild case as an example to illustrate the relationship between the number of intersections $n$ and the impact parameter $b$, as well as the photon trajectories around the Schwarzschild BH. In the left panel of Fig.~\ref{sch}, the black, orange, and red solid lines represent the direct emission, lensing ring, and photon ring, respectively. The right panel shows the corresponding photon trajectories around the BH, where the blue dashed line indicates the equatorial plane of BH. The black disk and the black dashed circle denote the BH and the photon sphere, respectively. It is clear that the range of impact parameters for the direct emission region is larger than those for the lensing ring and photon ring regions.
\begin{figure*}[htpb]
	\centering
	\includegraphics[width=0.49\textwidth]{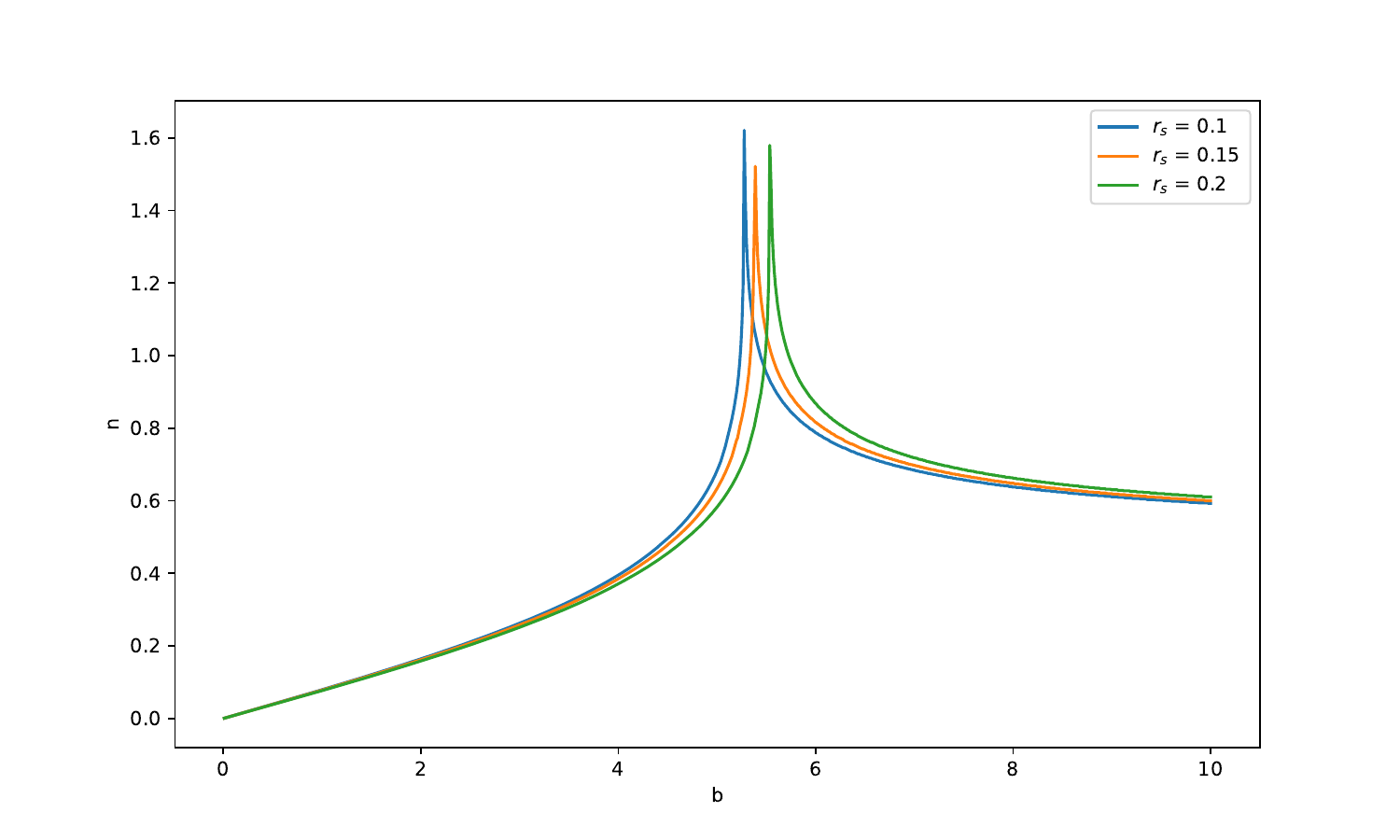}
	\includegraphics[width=0.49\textwidth]{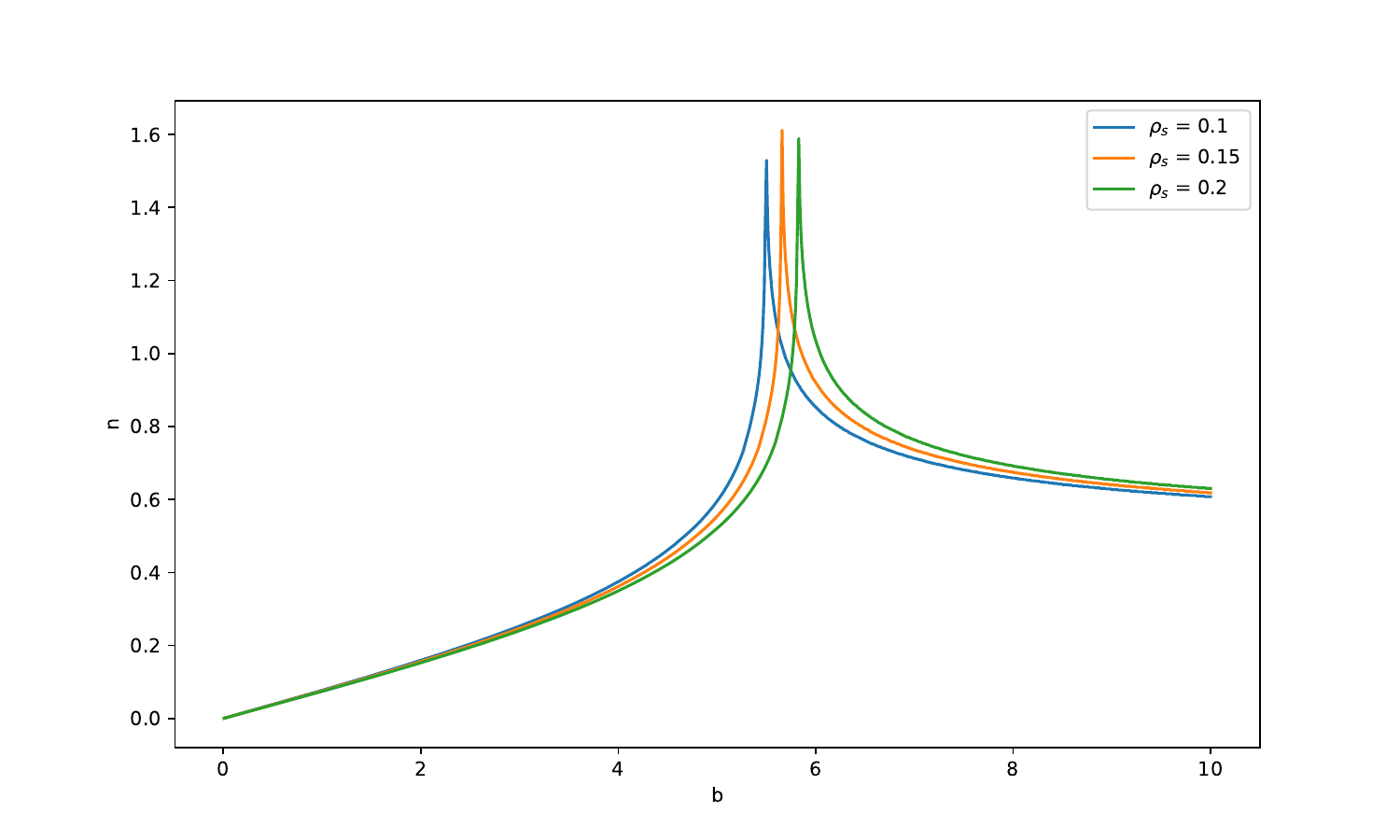}
	\caption{Variation of the intersection number with the impact parameter for the Schwarzschild-like BH under different parameter values. The left panel fixes $\rho_{s} = 0.01$ with $r_{s} = 0.1$, 0.15, and 0.2; the right panel fixes $r_{s} = 0.06$ with $\rho_{s} = 0.1$, 0.15, and 0.2.}
	\label{fig:nb_zong}	
\end{figure*}

Furthermore, in the left panel of Fig.~\ref{fig:nb_zong}, we show how the intersection number $n$ varies with the impact parameter $b$ for different values of $r_{s}$ with fixed $\rho_{s} = 0.01$, while in the right panel, we fix $r_{s} = 0.06$ and show the variation of $n$ with $b$ for different values of $\rho_{s}$. It can be found that when the value of $\rho_{s}$ is fixed, as the parameter $r_{s}$ increases, the corresponding curves as a whole show a trend of shifting to the right. This phenomenon illustrates that as the DM halo radius $r_{s}$ increases, the impact parameter $b$ associated with the lensing and photon rings also increases. Similarly, when $r_{s}$ takes a fixed value, the impact parameter of the system also shows a variation rule of increasing monotonically with the increase of $\rho_{s}$.

\begin{figure*}[htb]
	\centering
	\begin{subfigure}{0.32\textwidth}
		\includegraphics[height=4.6cm, keepaspectratio]{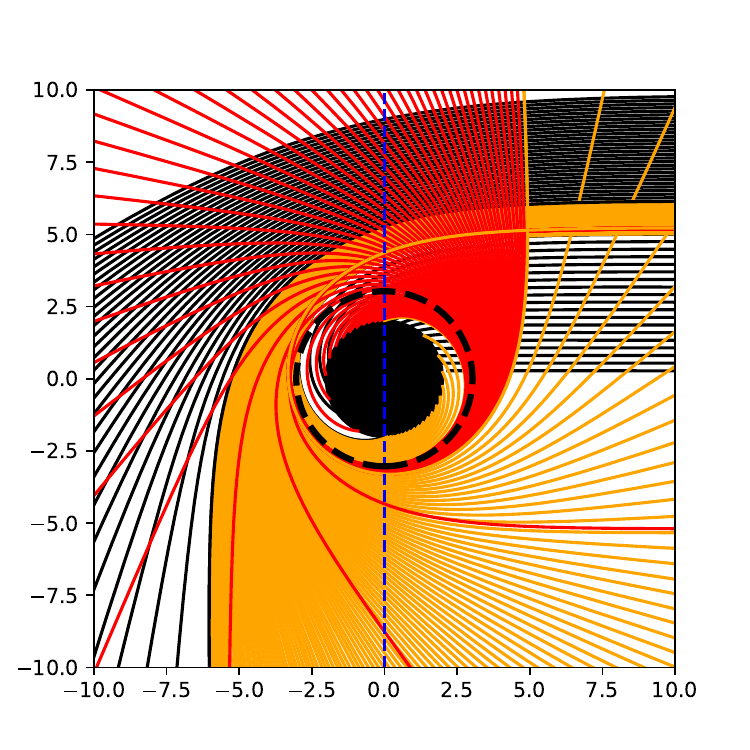} 
		\caption{$\rho_{s}=0.01$,
			$r_{s}=0.1$}
	\end{subfigure}
	\begin{subfigure}{0.32\textwidth}
		\includegraphics[height=4.6cm, keepaspectratio]{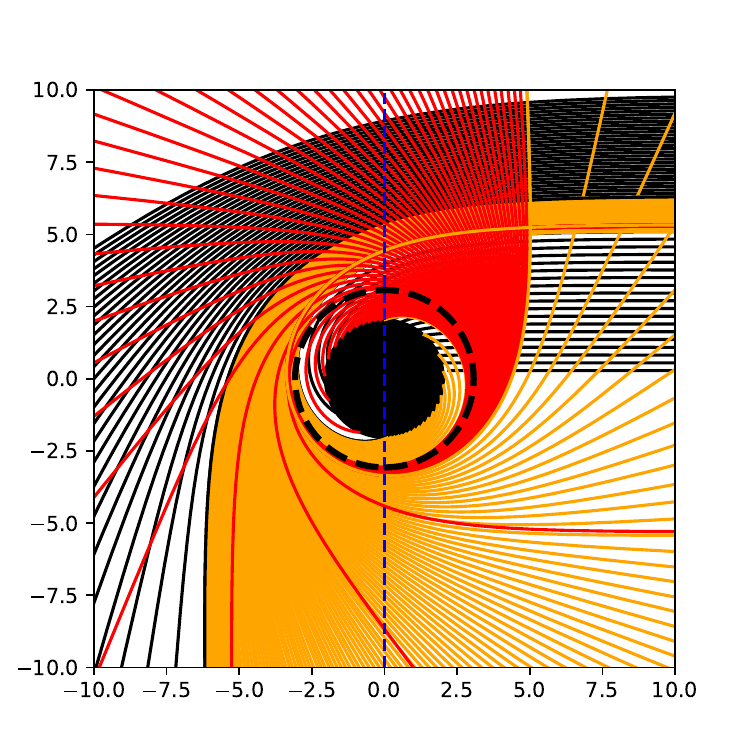} 
		\caption{$\rho_{s}=0.01$,
			$r_{s}=0.15$}
	\end{subfigure}
	\begin{subfigure}{0.32\textwidth}
		\includegraphics[height=4.6cm, keepaspectratio]{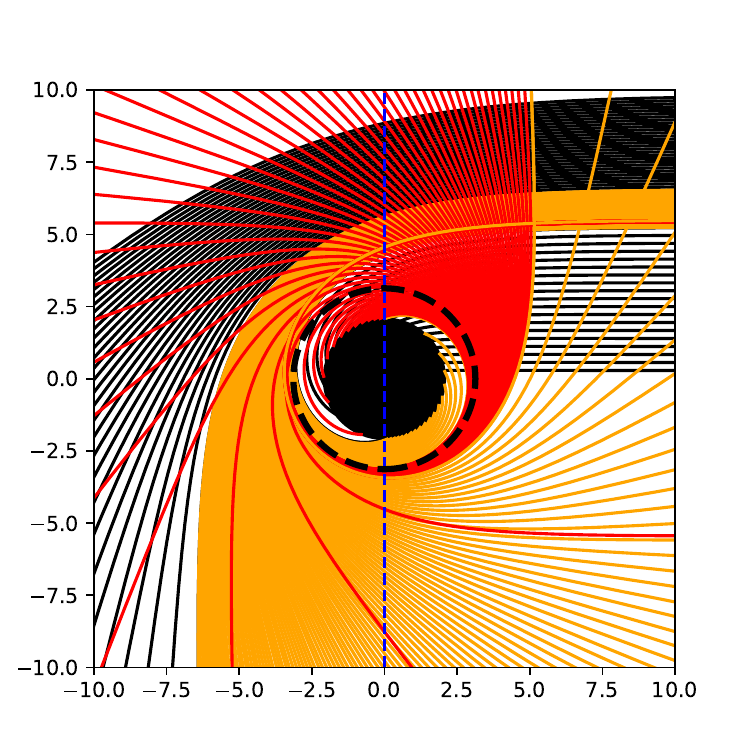} 
		\caption{$\rho_{s}=0.01$,
			$r_{s}=0.2$}
	\end{subfigure}
	\begin{subfigure}{0.32\textwidth}
		\includegraphics[height=4.6cm, keepaspectratio]{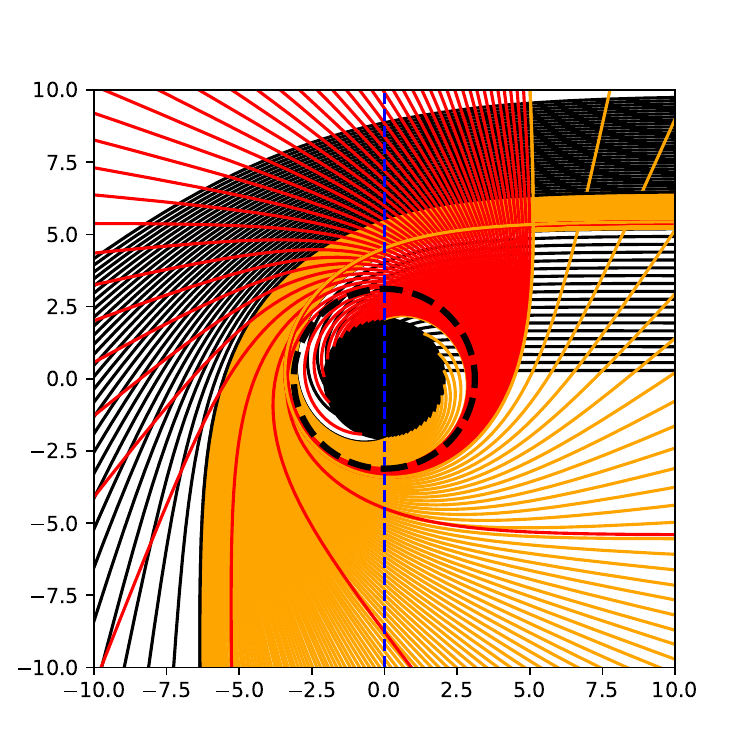} 
		\caption{$r_{s}=0.06$,
			$\rho_{s}=0.1$}
	\end{subfigure}
	\begin{subfigure}{0.32\textwidth}
		\includegraphics[height=4.6cm, keepaspectratio]{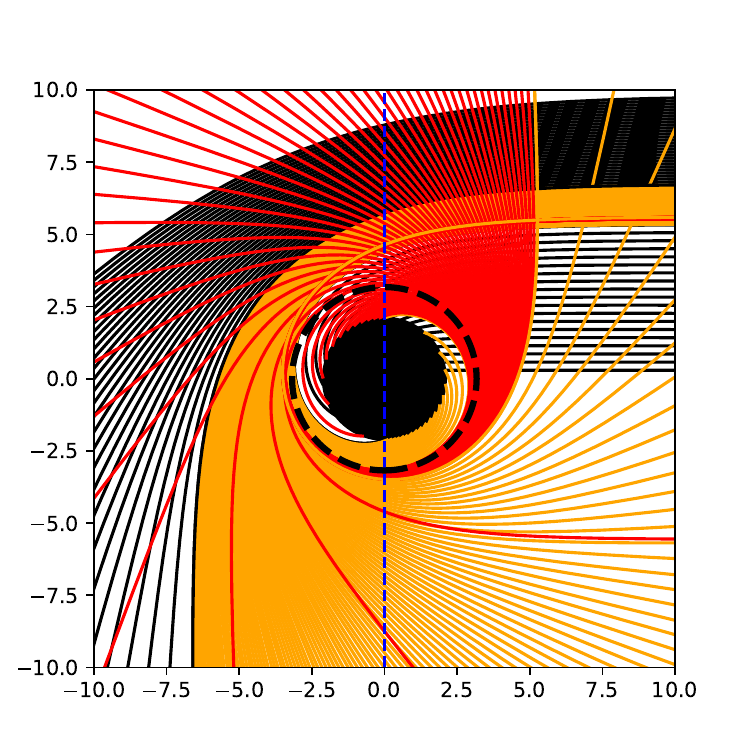} 
		\caption{$r_{s}=0.06$,
			$\rho_{s}=0.15$}
	\end{subfigure}
	\begin{subfigure}{0.32\textwidth}
		\includegraphics[height=4.6cm, keepaspectratio]{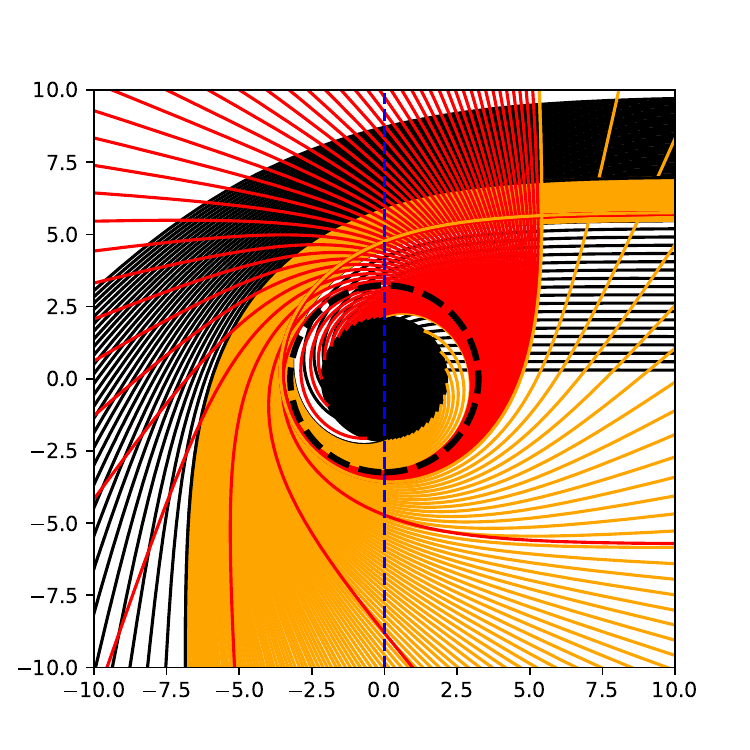} 
		\caption{$r_{s}=0.06$,
			$\rho_{s}=0.2$}
	\end{subfigure}
	\caption{The motion trajectories of photons around the Schwarzschild-like BH when $\rho_{s}$ and $r_{s}$ take different values.}
	\label{fig:photon}
\end{figure*}

Subsequently, we present in Fig.~\ref{fig:photon} the photon trajectories near the Schwarzschild-like BH corresponding to the parameter values used in Fig.~\ref{fig:nb_zong}. It can be clearly seen that, for fixed $\rho_{s}$, as $r_{s}$ increases, the yellow and red regions formed by the photon trajectories move farther away from the BH and become larger. This indicates that the scopes of the impact parameter corresponding to the lensing ring $b\in(b_2^-,b_3^-)\cup(b_3^+,b_2^+)$ and photon ring $b\in(b_3^-,b_3^+)$ also increase, which is consistent with the above discussion. A similar trend is also observed when $r_{s}$ is fixed and $\rho_{s}$ gradually increases.

Moreover, through further calculations, we have given the impact parameter intervals with  the number of intersections with the accretion disk. These interval values are presented in Tables \ref{tab11} and \ref{tab22}. The table also provides the calculations for various parameters, including the radius of the innermost stable circular orbit $r_{isco}$, the photon sphere radius $r_{ph}$, the event horizon radius $r_{h}$, and the critical impact parameter $b_{ph}$ .The radius of the innermost stable circular orbit can be determined using the following equation \cite{Wang:2023vcv}
\begin{equation}\label{19}
	\begin{split}
		r_{isco}=\frac{3f(r_{isco})f^{\prime}(r_{isco})}{2(f^{\prime}(r_{isco}))^{2}-f(r_{isco})f^{\prime\prime}(r_{isco})},
	\end{split}
\end{equation}
where the prime denotes differentiation with respect to the radial coordinate $r$. It is clear that in Tables ~\ref{tab11} and ~\ref{tab22}, all relevant physical quantities increase with the DM parameters. This is consistent with our previous discussion.
\begin{table*}[ht]
	\centering
	\renewcommand{\arraystretch}{1.6} 
	\begin{tabular}{p{2.3cm}p{1cm}p{1cm}p{1cm}p{1cm}p{1cm}p{1cm}p{1cm}p{1cm}p{1cm}}
		\hline\hline 
		$r_{s}$ &  ${b}_{1}^{-}$ &  ${b}_{2}^{-}$ & ${b}_{3}^{-}$ &${b}_{3}^{+}$ & ${b}_{2}^{+}$  & $r_{h}$ & $r_{ph}$ &  $b_{ph}$ &$r_{isco}$\\
		\hline
		Schwarzschild     & 2.848 & 5.015 & 5.188 & 5.226 & 6.128 & 2.000 & 3.000 & 5.196 &6.000 \\
		0.1               & 2.880 & 5.089 & 5.268 & 5.308 & 6.252 & 2.021 & 3.031 & 5.277 &6.064 \\
		0.15              & 2.922 & 5.184 & 5.373 & 5.416 & 6.416 & 2.048 & 3.072 & 5.382 &6.150 \\
		0.2               & 2.984 & 5.326 & 5.527 & 5.574 & 6.661 & 2.088 & 3.131 & 5.537 &6.280 \\
		\hline\hline 
	\end{tabular}
	\caption{The relevant physical quantities of the Schwarzschild BH and those of the Schwarzschild-like BH with $\rho_s = 0.01$ and $r_s = 0.1$, $0.15$, and $0.2$.}
	\label{tab11}
\end{table*}

\begin{table*}[ht]
	\centering
	\renewcommand{\arraystretch}{1.6} 
	\begin{tabular}{p{2.3cm}p{1cm}p{1cm}p{1cm}p{1cm}p{1cm}p{1cm}p{1cm}p{1cm}p{1cm}}
		\hline\hline 
		$\rho_{s}$ &  ${b}_{1}^{-}$ &  ${b}_{2}^{-}$ & ${b}_{3}^{-}$ &${b}_{3}^{+}$ & ${b}_{2}^{+}$  & $r_{h}$ & $r_{ph}$ &  $b_{ph}$ &$r_{isco}$\\
		\hline
		Schwarzschild     & 2.848 & 5.015 & 5.188 & 5.226 & 6.128 & 2.000 & 3.000 & 5.196 &6.000 \\
		0.1   & 2.966 & 5.287 & 5.484 & 5.531 & 6.596 & 2.076 & 3.114 & 5.495 &6.233 \\
		0.15  & 3.029 & 5.432 & 5.643 & 5.695 & 6.855 & 2.117 & 3.175 & 5.655 &6.356 \\
		0.2   & 3.092 & 5.584 & 5.809 & 5.867 & 7.134 & 2.158 & 3.238 & 5.822 &6.485 \\
		\hline\hline 
	\end{tabular}
	\caption{The relevant physical quantities of the Schwarzschild BH and those of the Schwarzschild-like BH for a fixed $r_{s} = 0.06$ and varying $\rho_{s} = 0.1$, $0.15$, and $0.2$.}
	\label{tab22}
\end{table*}

\subsection{\label{sec:level4.2}Transfer functions}
In the previous subsection, we obtained the photon trajectories around the Schwarzschild-like BH. Here, to facilitate the subsequent calculations, we introduce the transfer function to establish the connection between the photon’s impact parameter and the location of its intersection point on the accretion disk. 

\begin{figure*}[htb]
	\includegraphics[width=0.5\textwidth]{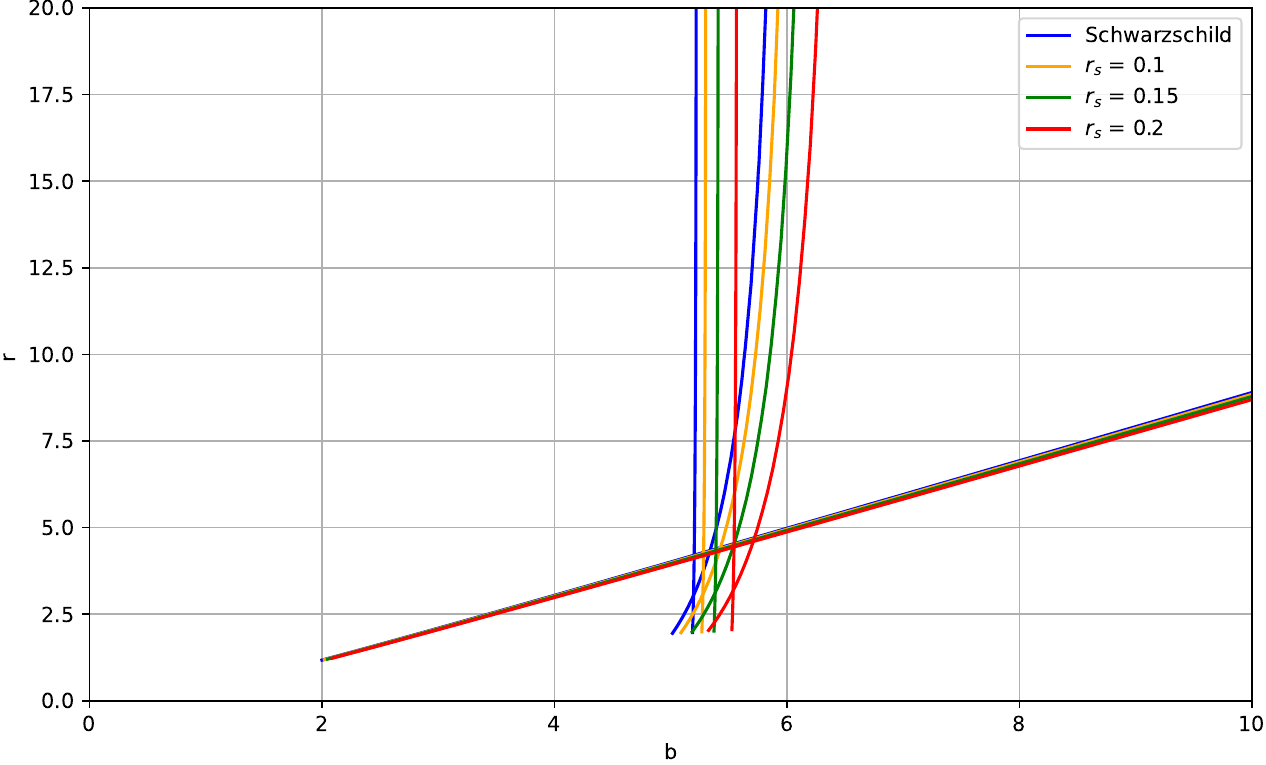}
	\includegraphics[width=0.5\textwidth]{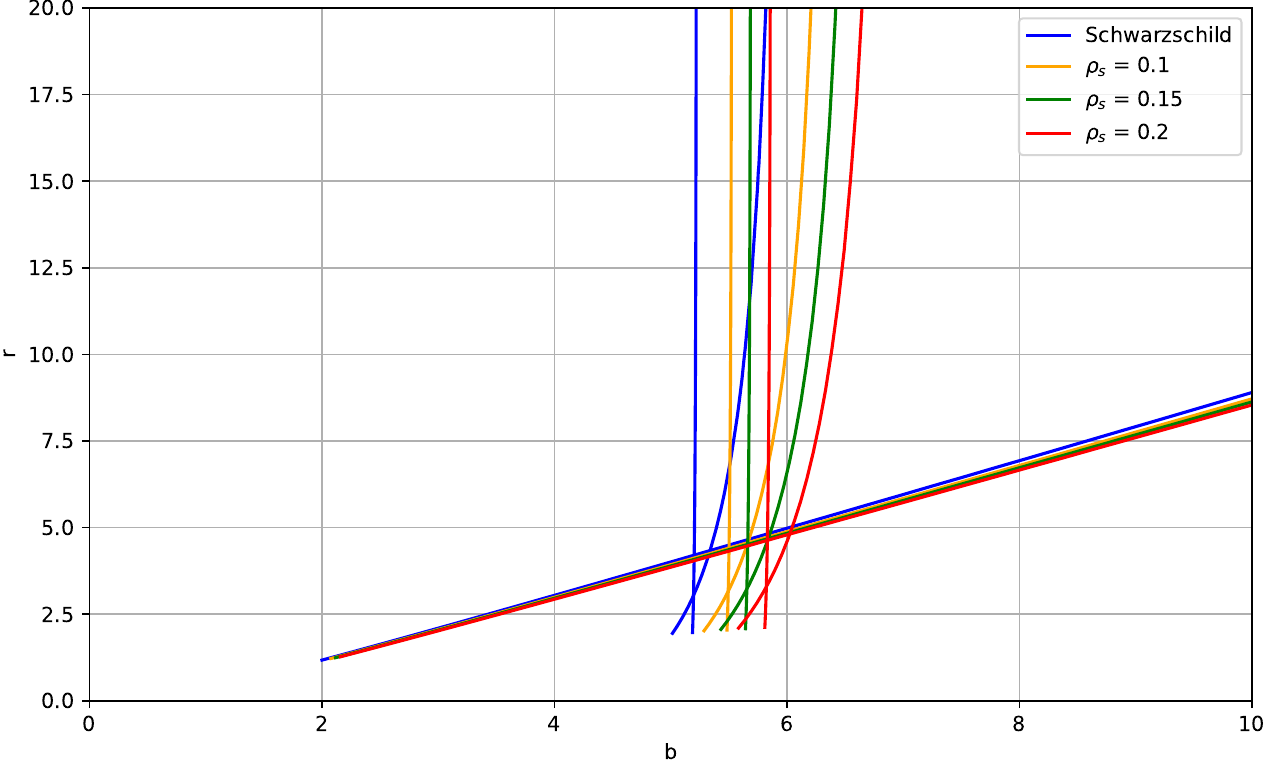}
	\caption{On the left side are the first three transfer functions corresponding to the cases where $\rho_{s}=0.01$, $r_{s}$ takes the values of 0.1, 0.15, 0.2, as well as the Schwarzschild case. On the right side are the first three transfer functions corresponding to the cases where $r_{s}=0.06$, $\rho_{s}$ takes the values of 0.1, 0.15, 0.2, as well as the Schwarzschild case.} 
	\label{fig:chuan}
\end{figure*}

We denote the transfer function as $r_m(b)$ ($m = 1, 2, 3, \cdots$), which represents the location of the $m$-th intersection point between the light ray and the accretion disk. The first three transfer functions can be formulated as what is shown below \cite{Tsukamoto:2021apr}
\begin{equation}\label{20}
	\begin{split}
		r_1(b) = \frac{1}{u\left(\frac{\pi}{2}, b\right)}, \quad b \in (b_1^-, +\infty)
	\end{split},
\end{equation}
\begin{equation}\label{21}
	\begin{split}
		r_2(b) = \frac{1}{u\left(\frac{3\pi}{2}, b\right)}, \quad b \in (b_2^-, b_2^+)
	\end{split},
\end{equation}
\begin{equation}\label{22}
	\begin{split}
		r_3(b) = \frac{1}{u\left(\frac{5\pi}{2}, b\right)}, \quad b \in (b_3^-, b_3^+)
	\end{split}.
\end{equation}
Here,  $u(\phi, b)$ is the solution of the photon trajectory equation \eqref {16}. We show the specific behavior of the first three transfer functions for different values of the DM halo parameters $\rho_{s}$ and $r_{s}$ in Fig.~\ref{fig:chuan}. We find that increasing the DM halo parameters significantly affects the second and third transfer functions, causing their corresponding impact parameters to gradually increase with the DM halo parameters. In contrast, for the first transfer function, its initial value slightly increases, while its slope experiences a minor decrease.

Since the observer measures the average brightness proportional to the flux \cite{Gralla:2019xty}, the slope of the transfer function $\frac{dr_m}{db}$ represents the demagnification factor of the optical image. A larger slope corresponds to a narrower interval of the impact parameter, resulting in a smaller contribution to the total observed intensity of the BH. Theoretically, photons in the photon ring region can intersect the accretion disk many times, but due to the steep slope of their transfer functions, their contribution to the observed intensity is almost negligible. Therefore, we only consider the first three transfer functions here.

\subsection{ Optical appearance of Schwarzschild-like black hole with thin accretion disk}
As discussed in the previous subsection, we consider an optically thin and geometrically thin, static accretion disk in the equatorial plane of a Schwarzschild-like BH as the sole radiation source, with isotropic emission. Therefore, the specific intensity of radiation emitted at frequency $\nu_e$, as measured by a distant observer at the north pole, is given by \cite{Wang:2023vcv}:
\begin{equation}\label{23}
	\begin{split}
		I_0(r,\nu_0)=g^3I_e(r,\nu_e)
	\end{split},
\end{equation}
where $g = \frac{\nu_0}{\nu_e} = \sqrt{f(r)}$ is the redshift factor, with $\nu_e$ and $\nu_0$ denoting the emitted and received frequencies of the radiation, respectively. Moreover, $I_0(r,\nu_0)$ represents the observed specific intensity, while $I_e(r,\nu_e)$ denotes the emitted specific intensity from the accretion disk, which depends solely on the radial coordinate. The total observed intensity $I_{obs}(r)$ can be obtained by integrating the specific intensity $I_0(r,\nu_0)$ over all frequencies
\begin{equation}\label{24}
	\begin{split}
		I_{obs}(r) = \int I_0 (r, \nu_0) d\nu_0 = \int g^4 I_e (r, \nu_e) d\nu_e = f(r)^2 I_{em}(r)
	\end{split}.
\end{equation}
Among them, we denote $I_{em}(r) = \int I_e (r, \nu_e) d\nu_e$ as the total emission intensity. From the above discussion, When photon trajectories are traced backward from the observer, they may intersect the accretion disk multiple times, as can be observed. Each intersection contributes additional intensity from the disk, therefore the total observed intensity at the observer is the sum of intensities from all these intersection points \cite{Peng:2020wun,Gralla:2019xty}
\begin{equation}\label{25}
	\begin{split}
		I_{obs}(b) = \sum_m f(r)^2 I_{em}(r) \Big|_{r = r_m(b)}
	\end{split}.
\end{equation}
Here, $r_m(b)$ is the transfer function. To observe the shadow and optical appearance of a Schwarzschild-like BH surrounded by a Dehnen-type DM halo, we consider projecting the optical appearance onto a two-dimensional plane under different emission models. Since photon motion along the unstable photon sphere constitutes the boundary of the BH shadow, the shadow shape can be obtained by projecting it from the BH plane to the observer's plane with celestial coordinates $(x, y)$. Here we define the celestial coordinate system as \cite{Kumar:2018ple}
\begin{equation}\label{26}
	\begin{split}
		x = \lim_{r_0 \to \infty} \left(-r_0^2 \sin \theta_0 \left(\frac{d\phi}{dr}\right)_{(r_0,\theta_0)}\right)
	\end{split},
\end{equation}
\begin{equation}\label{27}
	\begin{split}
		y = \lim_{r_0 \to \infty} \left(r_0^2 \sin \theta_0 \left(\frac{d\theta}{dr}\right)_{(r_0,\theta_0)}\right)
	\end{split},
\end{equation}
where, $r_0$ represents the position of the observer located at infinity, while $\theta_0$ stands for the angular position of the observer with respect to the plane of the BH.

\begin{figure*}[htb]
	\centering
	\begin{subfigure}{0.38\textwidth}
		\includegraphics[height=3.8cm, keepaspectratio]{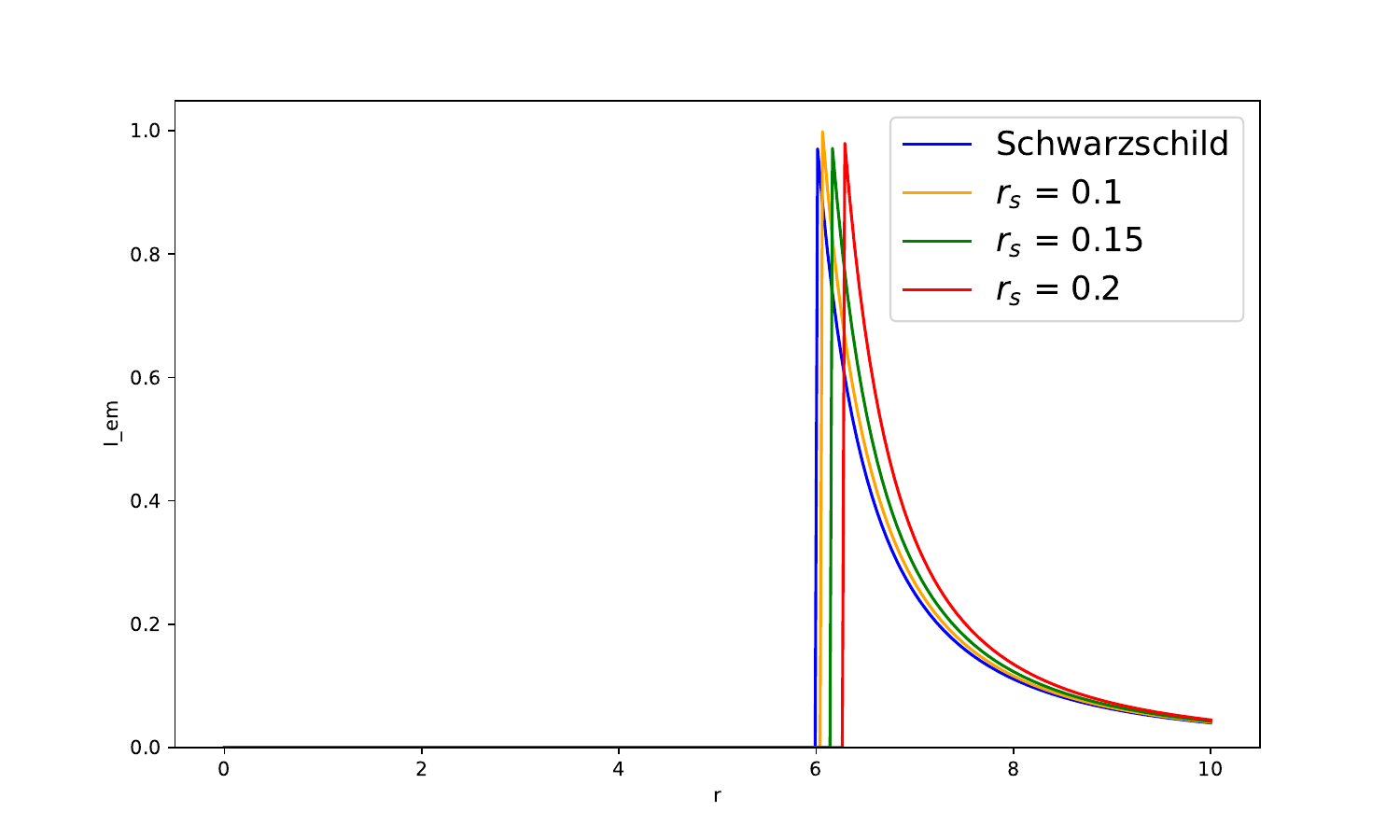} 
		\caption{}
	\end{subfigure}
	\begin{subfigure}{0.38\textwidth}
		\includegraphics[height=3.8cm, keepaspectratio]{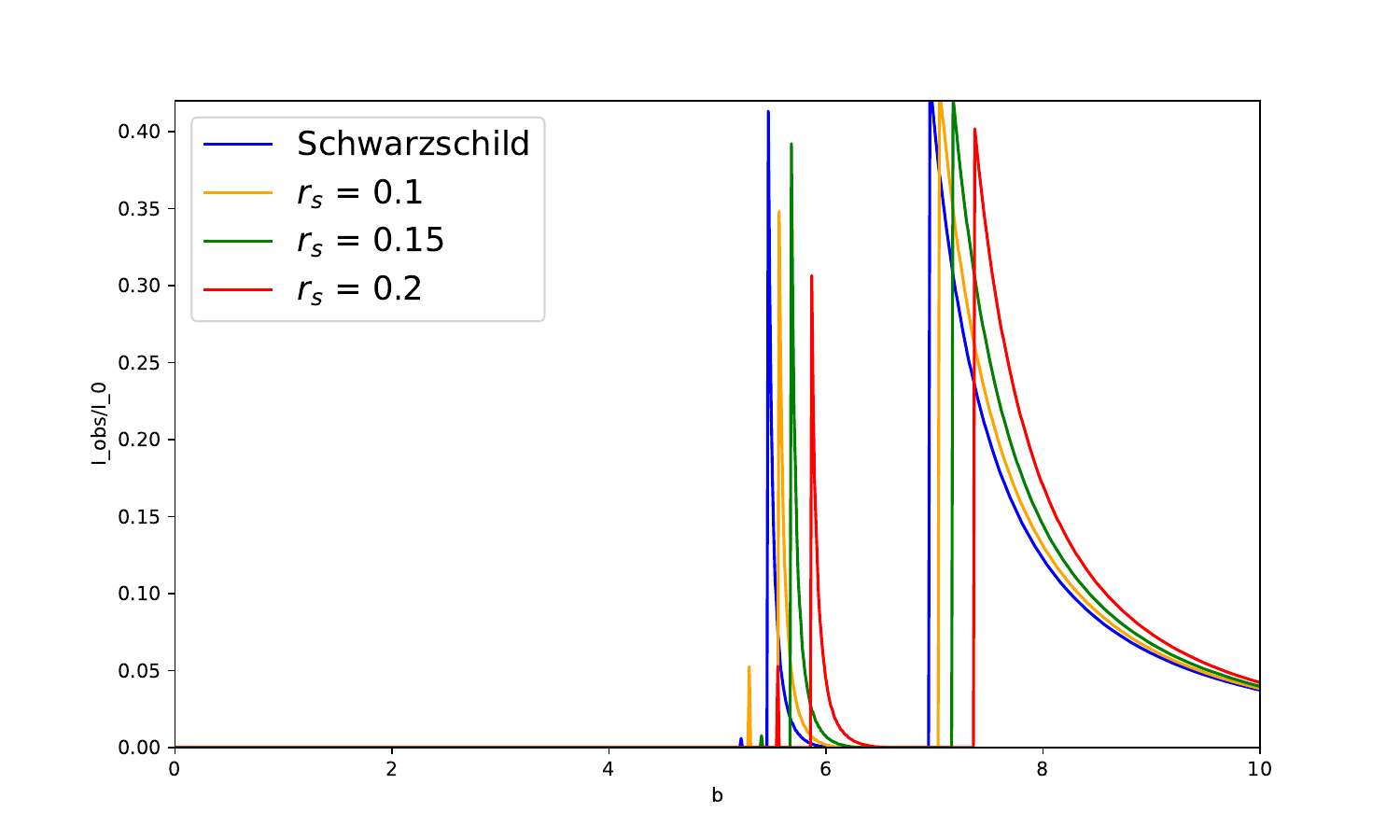} 
		\caption{}
	\end{subfigure}
	\begin{subfigure}{0.22\textwidth}
		\includegraphics[height=3.4cm, keepaspectratio]{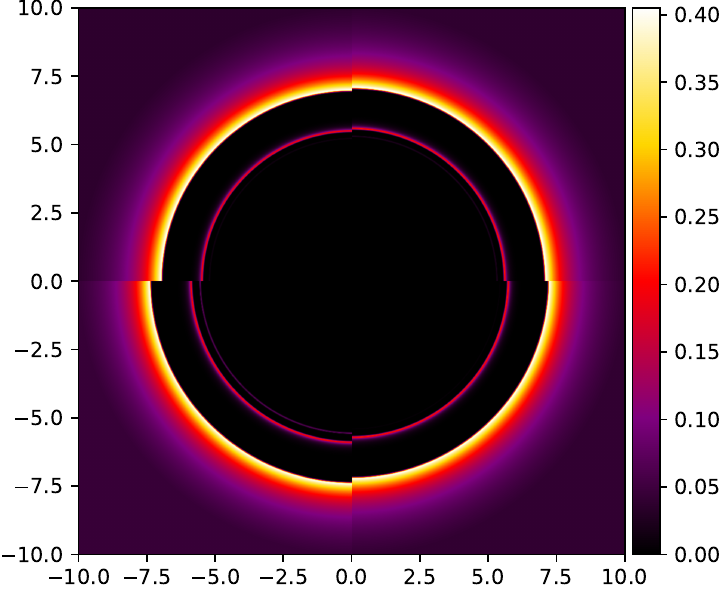} 
		\caption{}
	\end{subfigure}
	\begin{subfigure}{0.38\textwidth}
		\includegraphics[height=3.8cm, keepaspectratio]{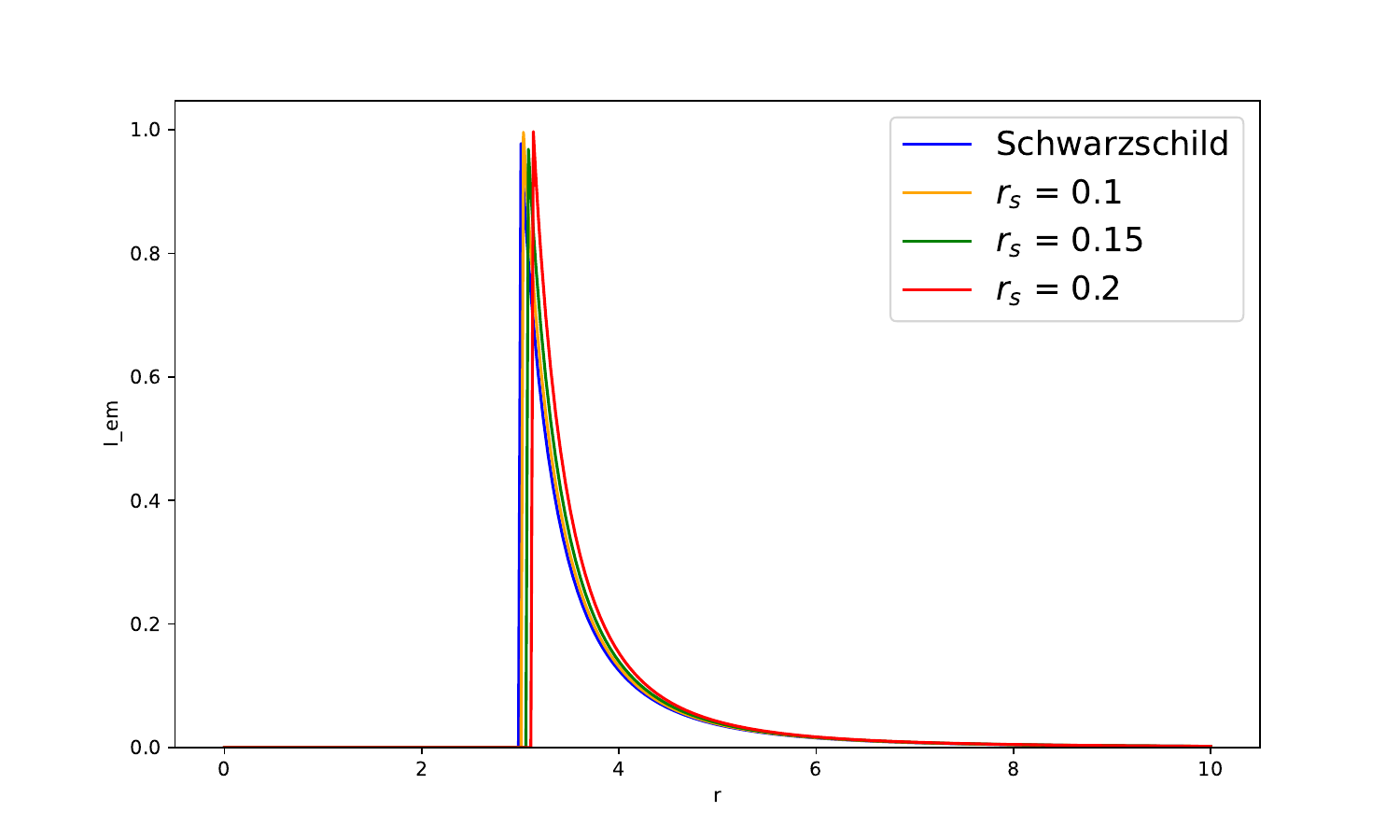} 
		\caption{}
	\end{subfigure}
	\begin{subfigure}{0.38\textwidth}
		\includegraphics[height=3.8cm, keepaspectratio]{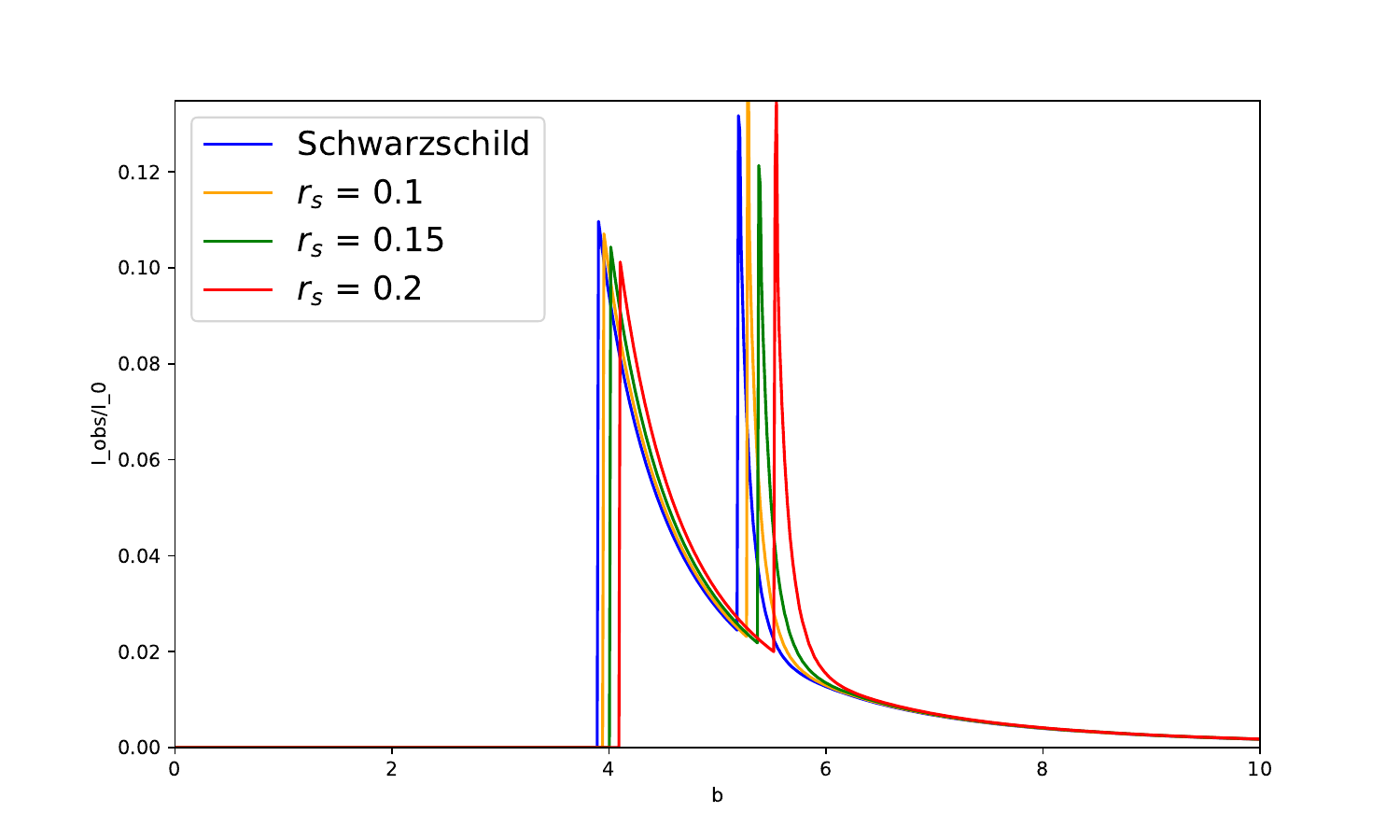} 
		\caption{}
	\end{subfigure}
	\begin{subfigure}{0.22\textwidth}
		\includegraphics[height=3.4cm, keepaspectratio]{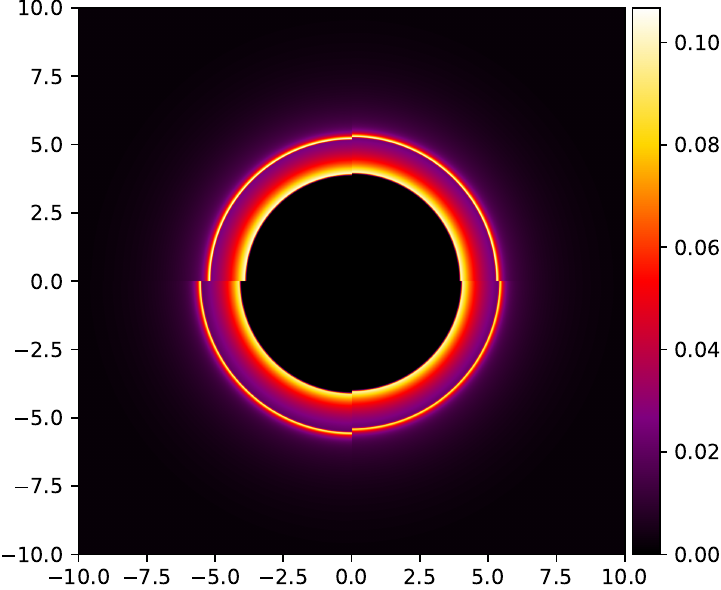} 
		\caption{}
	\end{subfigure}
	\begin{subfigure}{0.38\textwidth}
		\includegraphics[height=3.8cm, keepaspectratio]{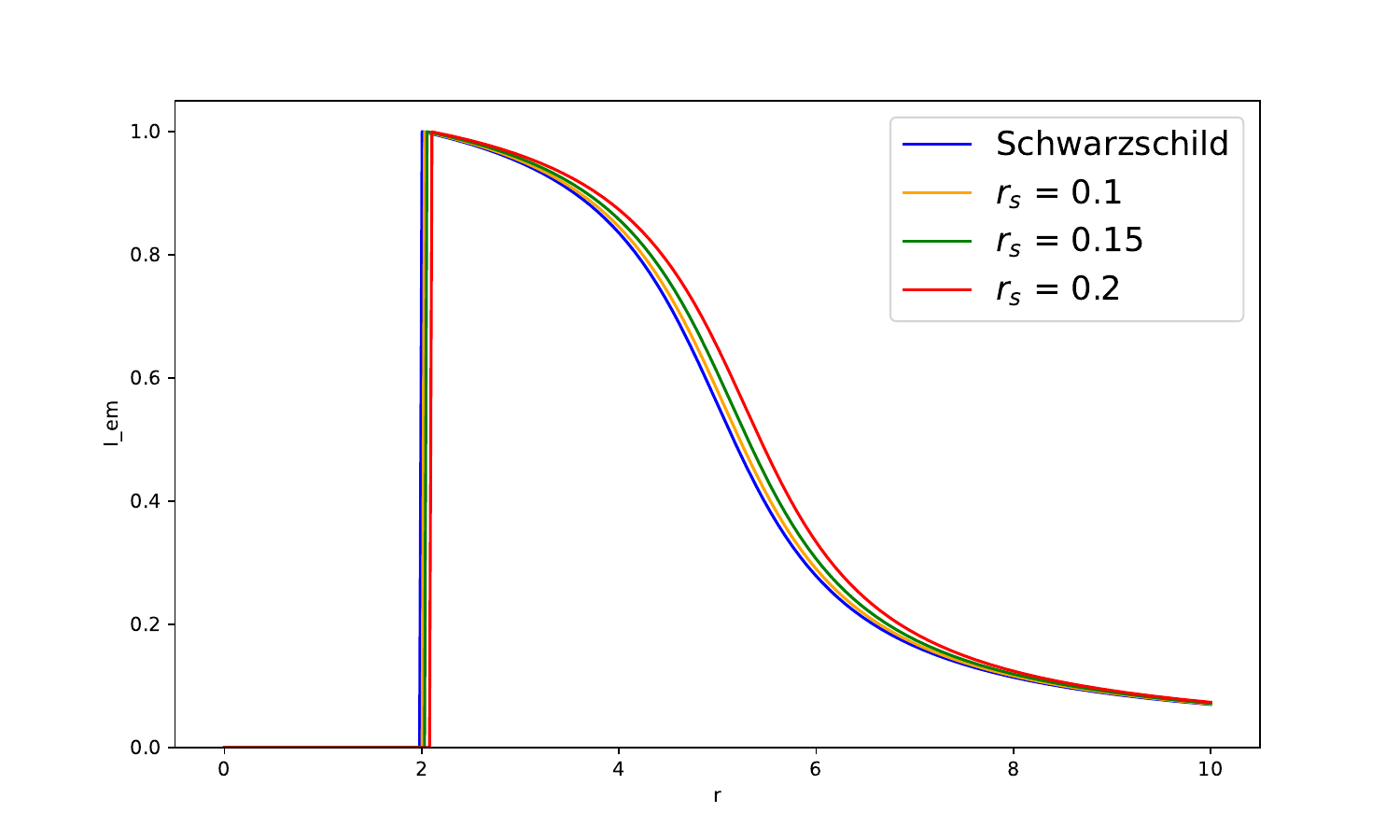} 
		\caption{}
	\end{subfigure}
	\begin{subfigure}{0.38\textwidth}
		\includegraphics[height=3.8cm, keepaspectratio]{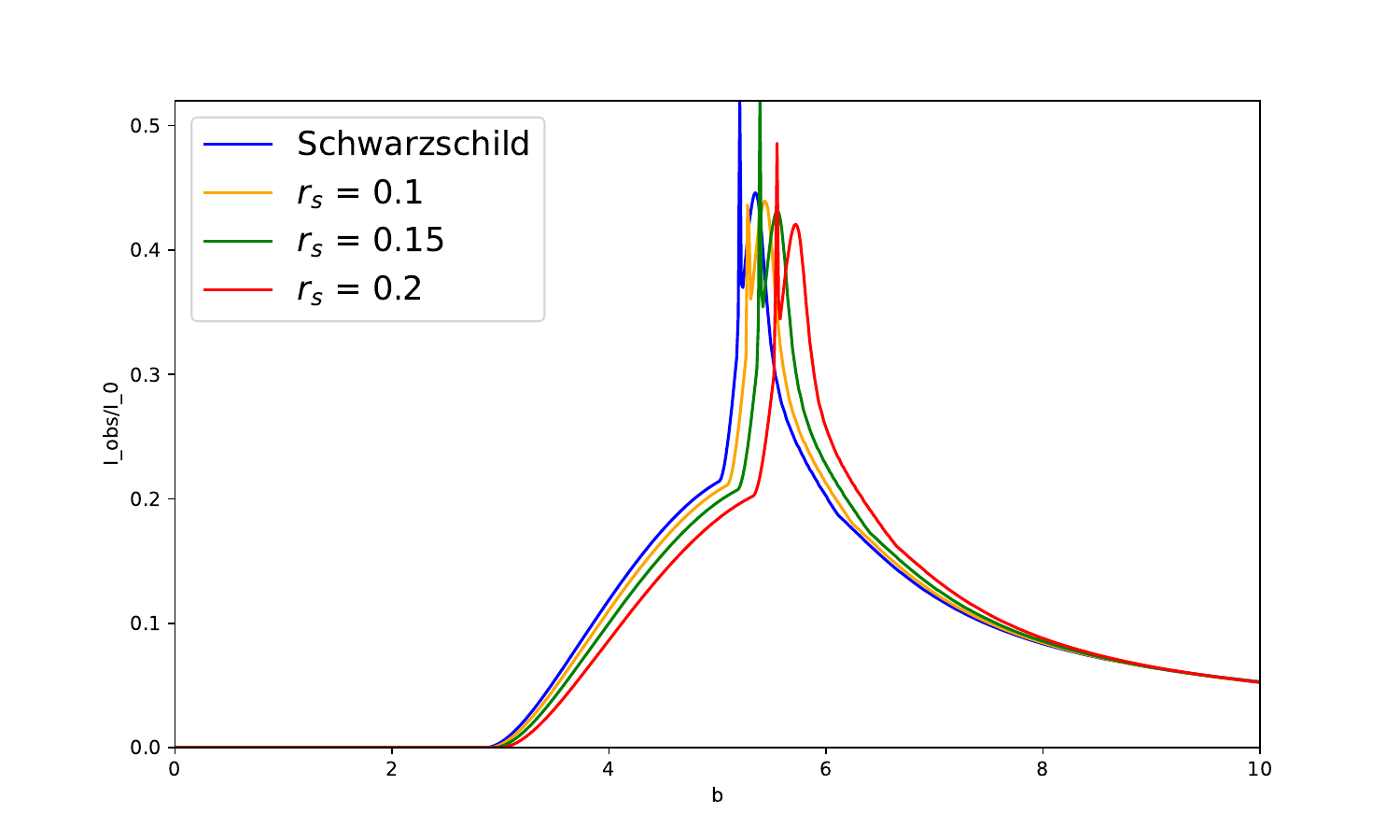} 
		\caption{}
	\end{subfigure}
	\begin{subfigure}{0.22\textwidth}
		\includegraphics[height=3.4cm, keepaspectratio]{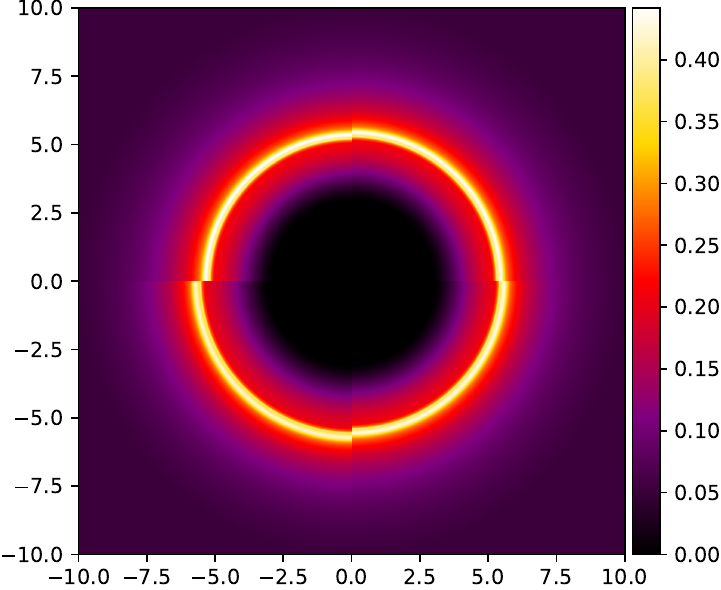} 
		\caption{}
	\end{subfigure}
	\caption{Emission intensity (left column), received intensity (middle column), and corresponding optical appearance (right column) for three models with fixed $\rho_{s} = 0.01$. In the left two columns, the blue, orange, green, and red curves represent the Schwarzschild case and the DM parameters with $r_{s} = 0.1$, 0.15, and 0.2, respectively. The right panel composites the optical appearances of: Schwarzschild BH (top left), the Schwarzschild-like BH with $r_{s} = 0.1$ (top right), $r_{s} = 0.15$ (bottom right), and $r_{s} = 0.2$ (bottom left).}
	\label{fig:rs}
\end{figure*}

\begin{figure*}[htb]
	\centering
	\begin{subfigure}{0.38\textwidth}
		\includegraphics[height=3.8cm, keepaspectratio]{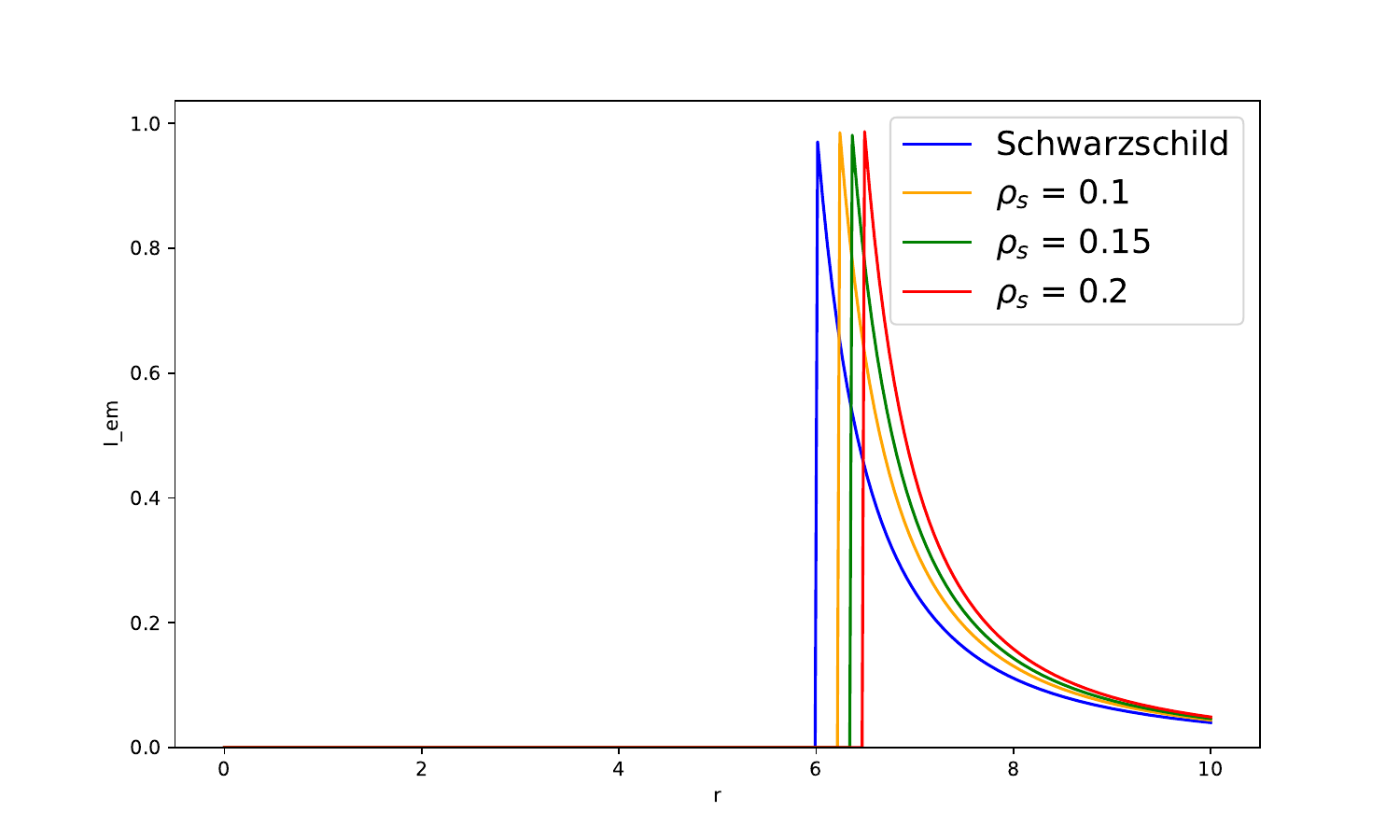} 
		\caption{}
	\end{subfigure}
	\begin{subfigure}{0.38\textwidth}
		\includegraphics[height=3.8cm, keepaspectratio]{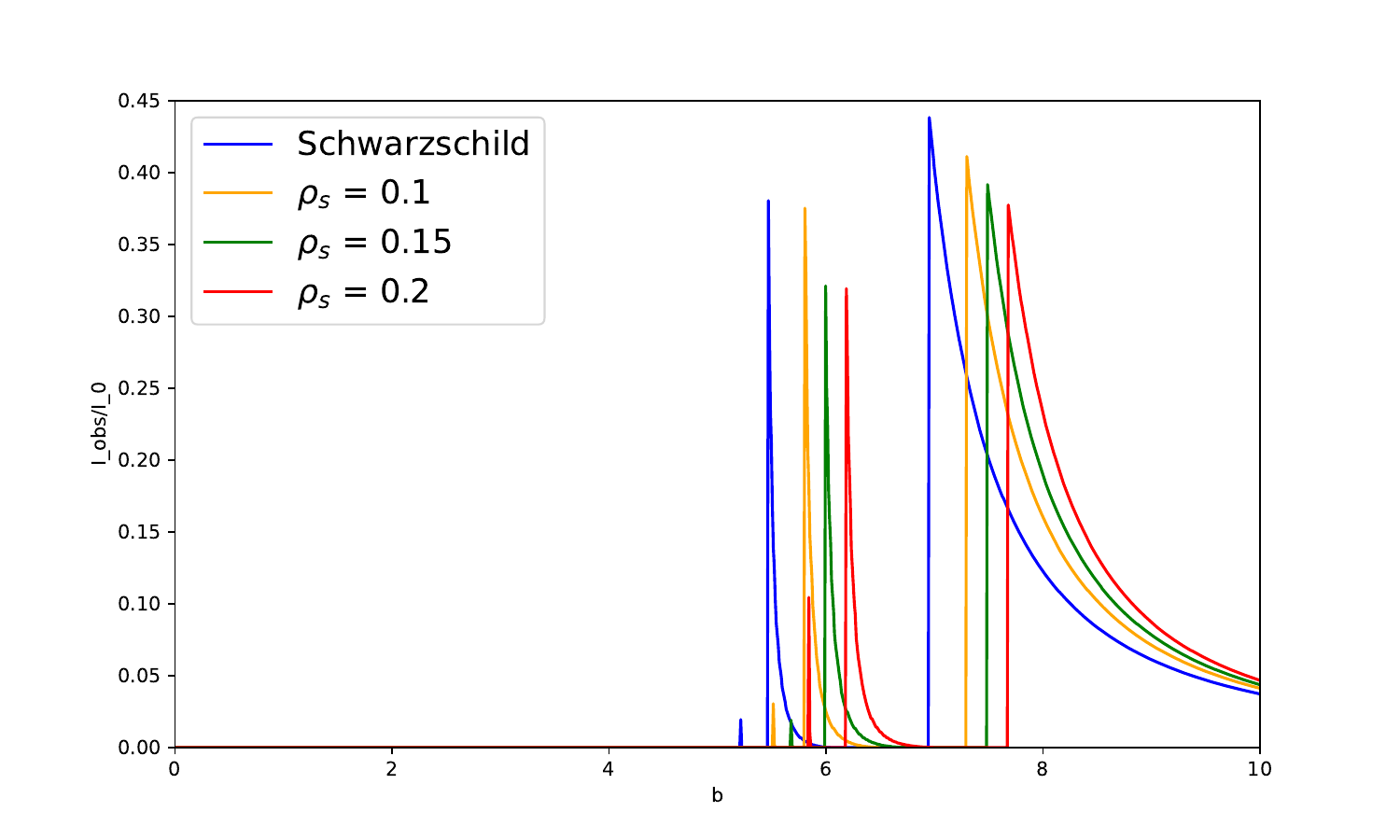} 
		\caption{}
	\end{subfigure}
	\begin{subfigure}{0.22\textwidth}
		\includegraphics[height=3.4cm, keepaspectratio]{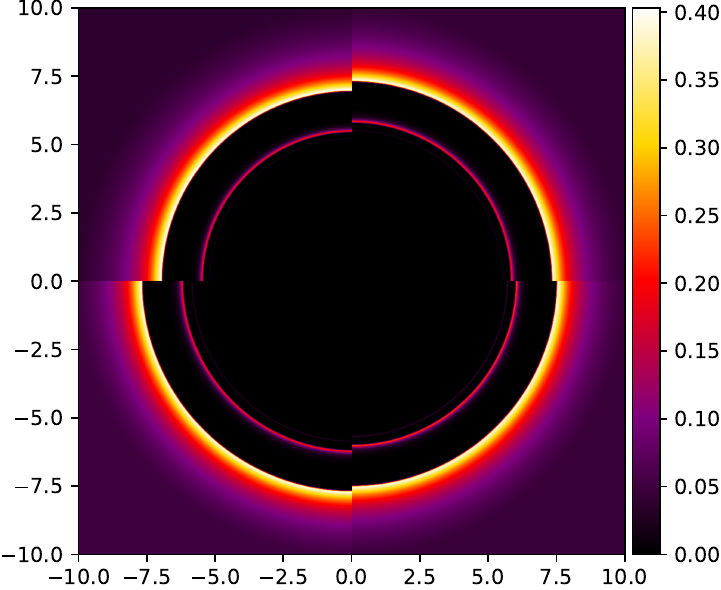} 
		\caption{}
	\end{subfigure}
	\begin{subfigure}{0.38\textwidth}
		\includegraphics[height=3.8cm, keepaspectratio]{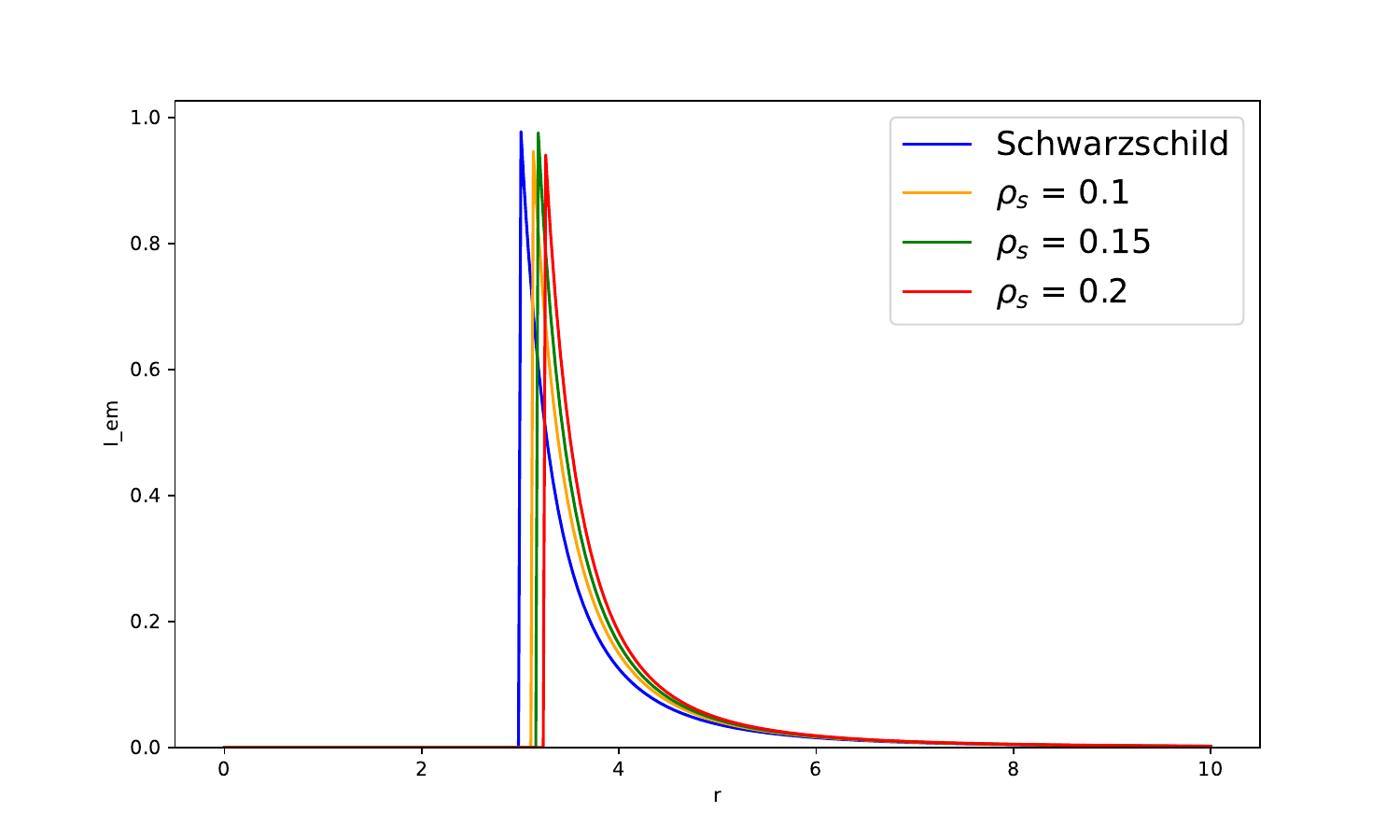} 
		\caption{}
	\end{subfigure}
	\begin{subfigure}{0.38\textwidth}
		\includegraphics[height=3.8cm, keepaspectratio]{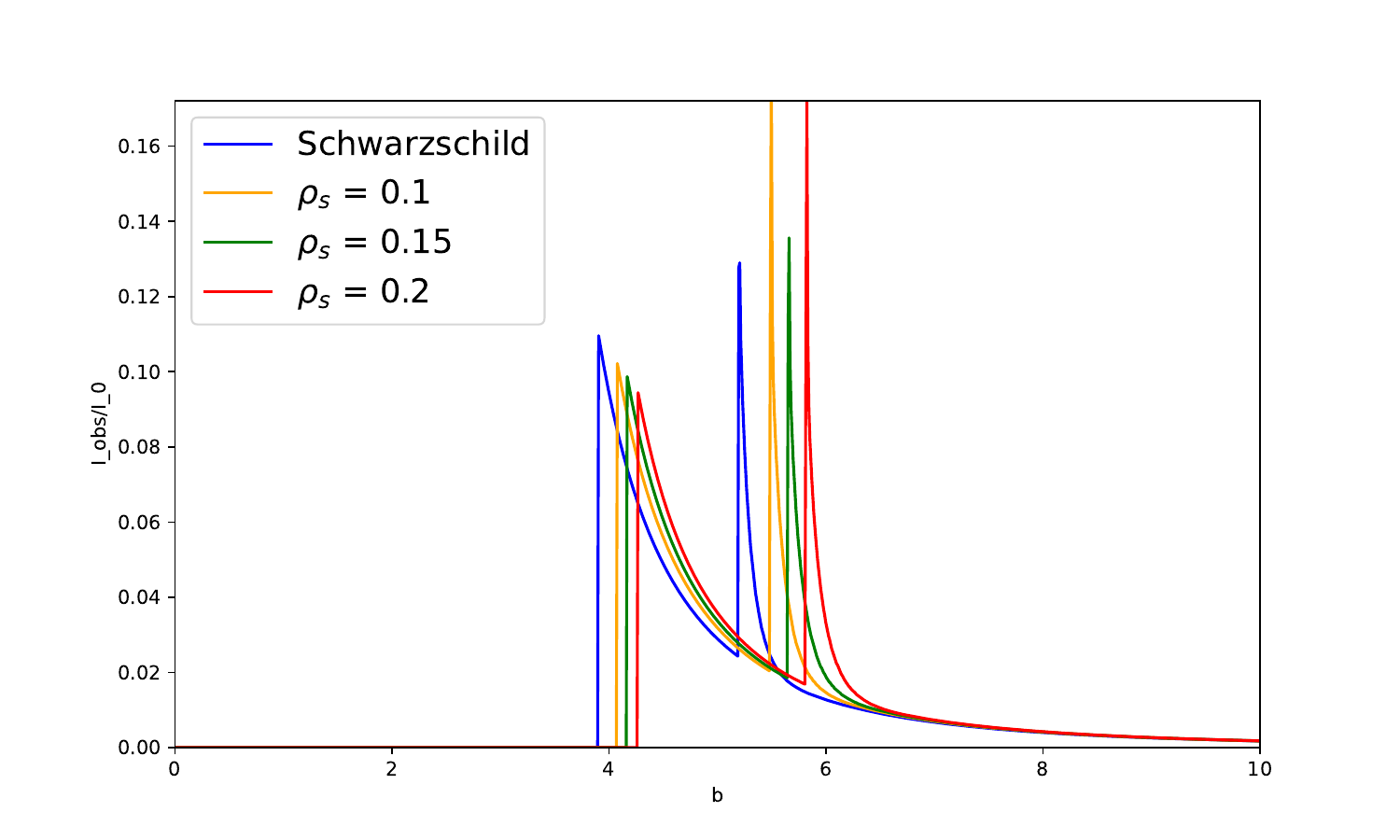} 
		\caption{}
	\end{subfigure}
	\begin{subfigure}{0.22\textwidth}
		\includegraphics[height=3.4cm, keepaspectratio]{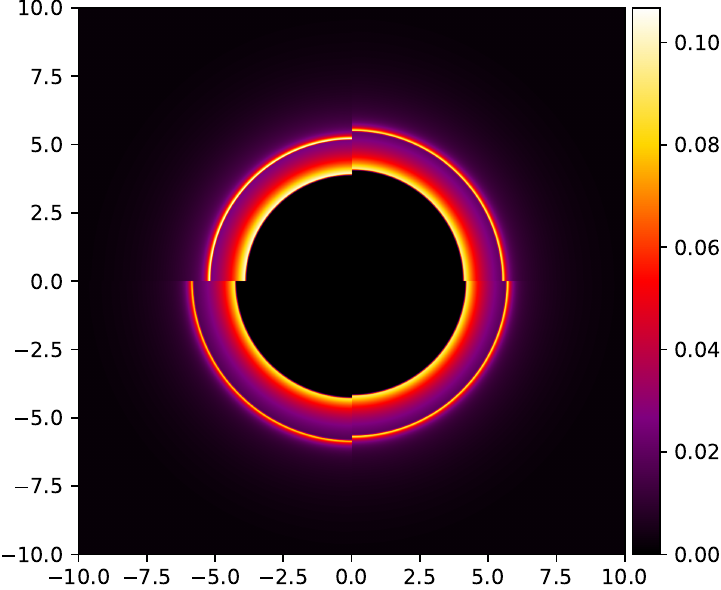} 
		\caption{}
	\end{subfigure}
	\begin{subfigure}{0.38\textwidth}
		\includegraphics[height=3.8cm, keepaspectratio]{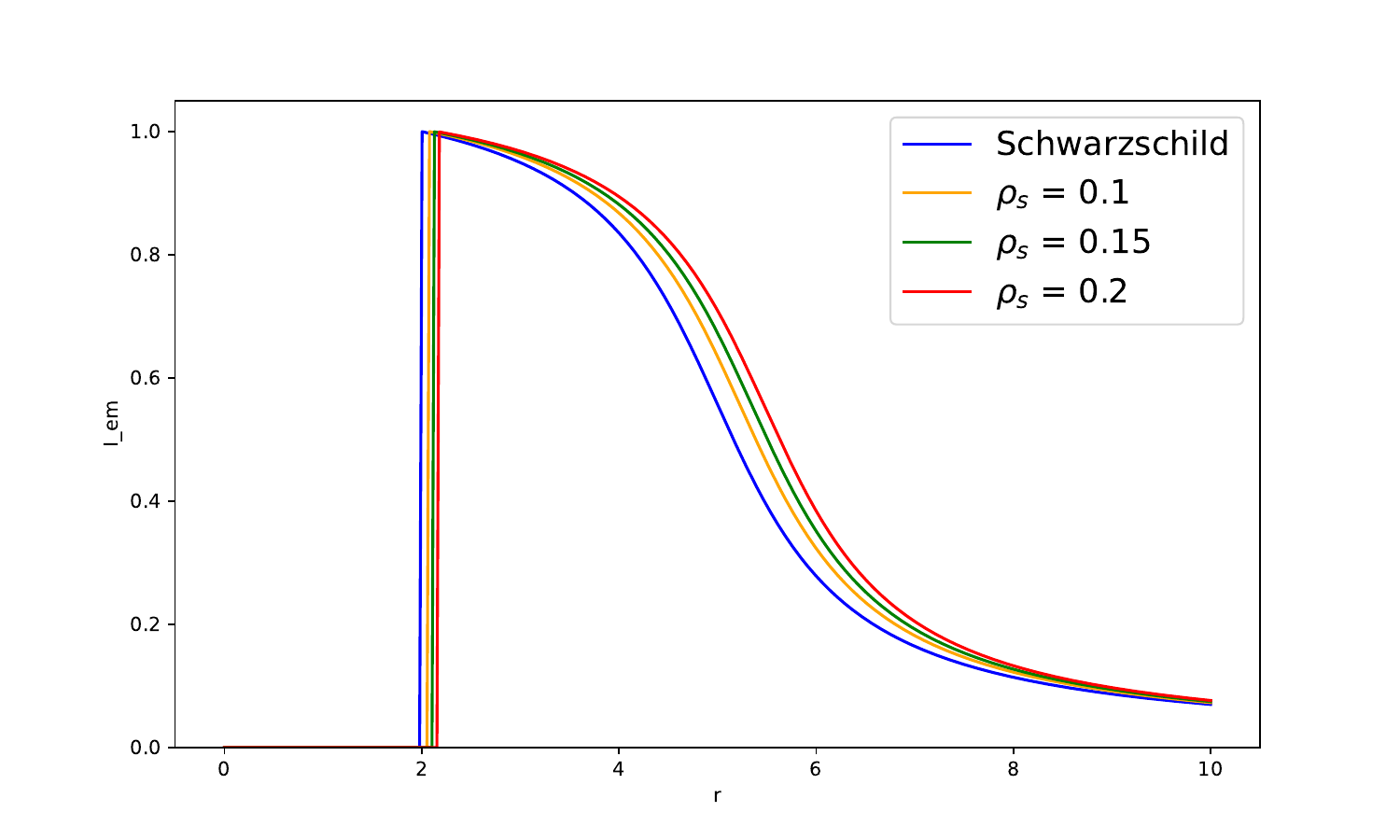} 
		\caption{}
	\end{subfigure}
	\begin{subfigure}{0.38\textwidth}
		\includegraphics[height=3.8cm, keepaspectratio]{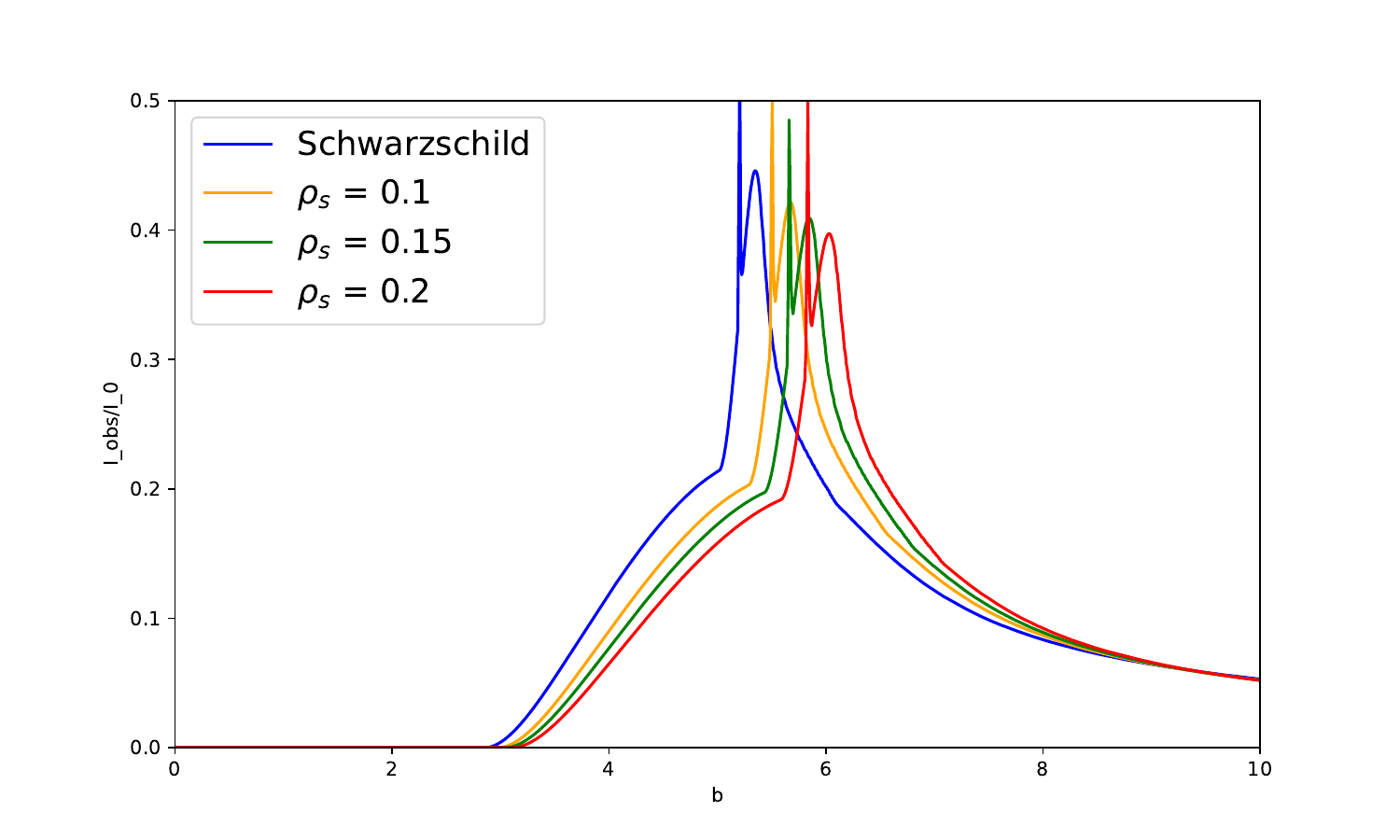} 
		\caption{}
	\end{subfigure}
	\begin{subfigure}{0.22\textwidth}
		\includegraphics[height=3.4cm, keepaspectratio]{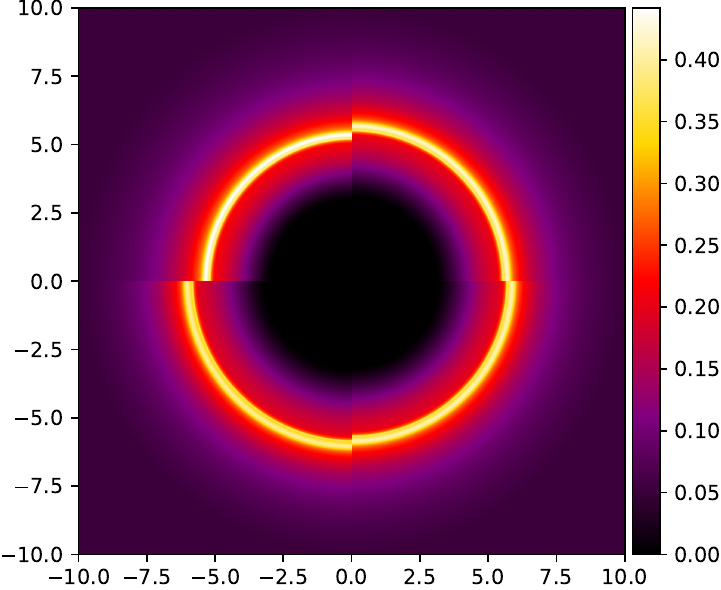} 
		\caption{}
	\end{subfigure}
	\caption{Emission intensity (left column), received intensity (middle column), and corresponding optical appearance (right column) for three models with fixed $r_{s} = 0.06$. In the left two columns, the blue, orange, green, and red curves represent the Schwarzschild case and the DM parameters with  $\rho_{s} = 0.1$, 0.15, and 0.2, respectively. The right panel composites the optical appearances of: Schwarzschild BH (top left), the Schwarzschild-like BH with $\rho_{s} = 0.1$ (top right), $\rho_{s} = 0.15$ (bottom right), and $\rho_{s} = 0.2$ (bottom left).}
	\label{fig:ros}
\end{figure*}

Therefore, it suffices for us to define the emission intensity of the accretion disk and calculate the corresponding received intensity using Eq.~\eqref {25}, and then project this intensity onto the two-dimensional plane in the celestial coordinate system to obtain the optical appearance of the BH.
We consider three static accretion disk models:
\begin{itemize}
	\item Model 1: The emission intensity $I_{em}(r)$ from the accretion disk experiences a rapid increase up to a peak at the innermost stable circular orbit $r_{isco}$, and then declines in accordance with a quadratic decay, namely,
	\begin{equation}\label{28}
		\begin{split}
			I_{em}(r)=\begin{cases}
				I_{0}\left[\frac{1}{r-(r_{isco}-1)}\right]^{2},&r > r_{isco},\\
				0,&r\leq r_{isco},
			\end{cases}
		\end{split}
	\end{equation}
	where $I_0$ is the maximum emission intensity, and $r_{isco}$ is the radius of the innermost stable circular orbit defined previously.  
	
	\item Model 2: In this model, we assume the emission intensity $I_{em}(r)$ of the accretion disk sharply reaches its peak at the photon sphere radius $r_{ph}$, and then decays in the form of a cubic power.
	\begin{equation}\label{29}
		\begin{split}	
			I_{em}(r)=\begin{cases}
				I_{0}\left[\frac{1}{r-(r_{ph}-1)}\right]^{3}, & r > r_{ph},\\ 			
				0, & r \leq r_{ph}.
			\end{cases}
		\end{split}
	\end{equation}
	
	\item Model 3: The emission intensity $I_{em}(r)$ of the accretion disk sharply increases to a peak value at the event horizon radius $r_h$, and then starts to decay slowly.
	\begin{equation}\label{30}
		\begin{split}	
			I_{em}(r)=\begin{cases}
				I_{0}\frac{\frac{\pi}{2}-\arctan\left[r-(r_{isco}-1)\right]}{\frac{\pi}{2}-\arctan\left[r_{h}-(r_{isco}-1)\right]}, & r > r_{h},\\
				0, & r \leq r_{h}.
			\end{cases}
		\end{split}
	\end{equation}
\end{itemize}
We then present the optical appearances of the Schwarzschild-like BHs with different DM halo parameters in Figs.~\ref{fig:rs} and \ref{fig:ros}. In these figures, the left two columns respectively show the emission and received intensities for the three accretion disk models under different parameter values, while the rightmost column displays their corresponding optical appearances projected onto the two-dimensional plane in the celestial coordinate.

In Fig. ~\ref{fig:rs}, the blue, orange, green, and red solid lines correspond to the Schwarzschild case and Schwarzschild-like cases with DM halo parameters fixed at $\rho_{s}=0.01$ for $r_{s}=0.1$, $0.15$, and $0.2$, respectively. We summarize these main characteristics as follows:
\begin{itemize}
	
	\item The total intensity received by the observer exhibits significant variations across different models. In the first model, three distinct sharp peaks are clearly identifiable. The first peak corresponds to the contribution from the photon ring region, which, as previously discussed, has a negligible impact on the BH’s optical appearance and is almost invisible in the figure (see panel (c) in Fig. ~\ref{fig:rs}). In contrast, the second model only displays two prominent sharp peaks, primarily contributed by direct emission and the lensing ring, with a minimal addition from the photon ring. Meanwhile, in the third model, the total observed intensity features two peaks. Due to their close proximity, these peaks manifest as a single broad bright ring in the image (see panel (i) in Fig. ~\ref{fig:rs}).
	
	\item As the DM halo radius $r_s$ gradually increases, both the emission intensity and the received intensity exhibit a rightward shift. This phenomenon suggests that several key physical quantities of the Schwarzschild-like BH surrounded by a Dehnen-type DM halo, such as the radius of the innermost stable circular orbit, the photon sphere radius, and the event horizon radius, also increase progressively with increasing $r_s$.
	
	\item As the radius of the DM halo $ r_s $ increases, we find that the optical appearance of the Schwarzschild-like BH in all three accretion disk models gradually deviates from that of a Schwarzschild BH, with the bright ring appearing farther away from the BH.  
\end{itemize}

Additionally, in Fig.~\ref{fig:ros}, we present the Schwarzschild case and Schwarzschild-like cases with the DM halo parameters fixed at $ r_s = 0.06 $ for $ \rho_s = 0.1$, $ 0.15$, and $ 0.2 $, respectively. The results are consistent with those in Fig.~\ref{fig:rs}, further demonstrating that the optical appearance of Schwarzschild-like BH embedded in a Dehnen-type DM halo exhibits significant deviations from the Schwarzschild BH case. This provides a potential way to distinguish Schwarzschild-like BHs surrounded by a Dehnen-type DM halo from their Schwarzschild counterparts.

\section{\label{sec:level5}The images of a Schwarzschild-like black hole with thin spherical accretions}
In previous section, the thin accretion disk considered in this study represents a simplified theoretical model. It is based on the assumption that after matter existing in the universe gets captured by a BH, it gradually forms the shape of a thin accretion disk. However, when the accreting matter possesses relatively low angular momentum, the scenario differs fundamentally. In this case, instead of forming a thin accretion disk, the matter will accrete radially onto the BH, ultimately establishing a spherical accretion state \cite{Yuan:2014gma}. Building upon this framework, we investigate the optical appearance of a Schwarzschild-like BH embedded in a Dehnen-type DM halo under spherical accretion. Specifically, we examine how variations in the DM halo's density $\rho_{s}$ and radius $r_s$ influence the observational features.

\subsection{\label{sec:level5.1}Optical appearance of Schwarzschild-like black hole with static thin spherical accretions}
\begin{figure*}[htb]
	\includegraphics[width=0.46\textwidth]{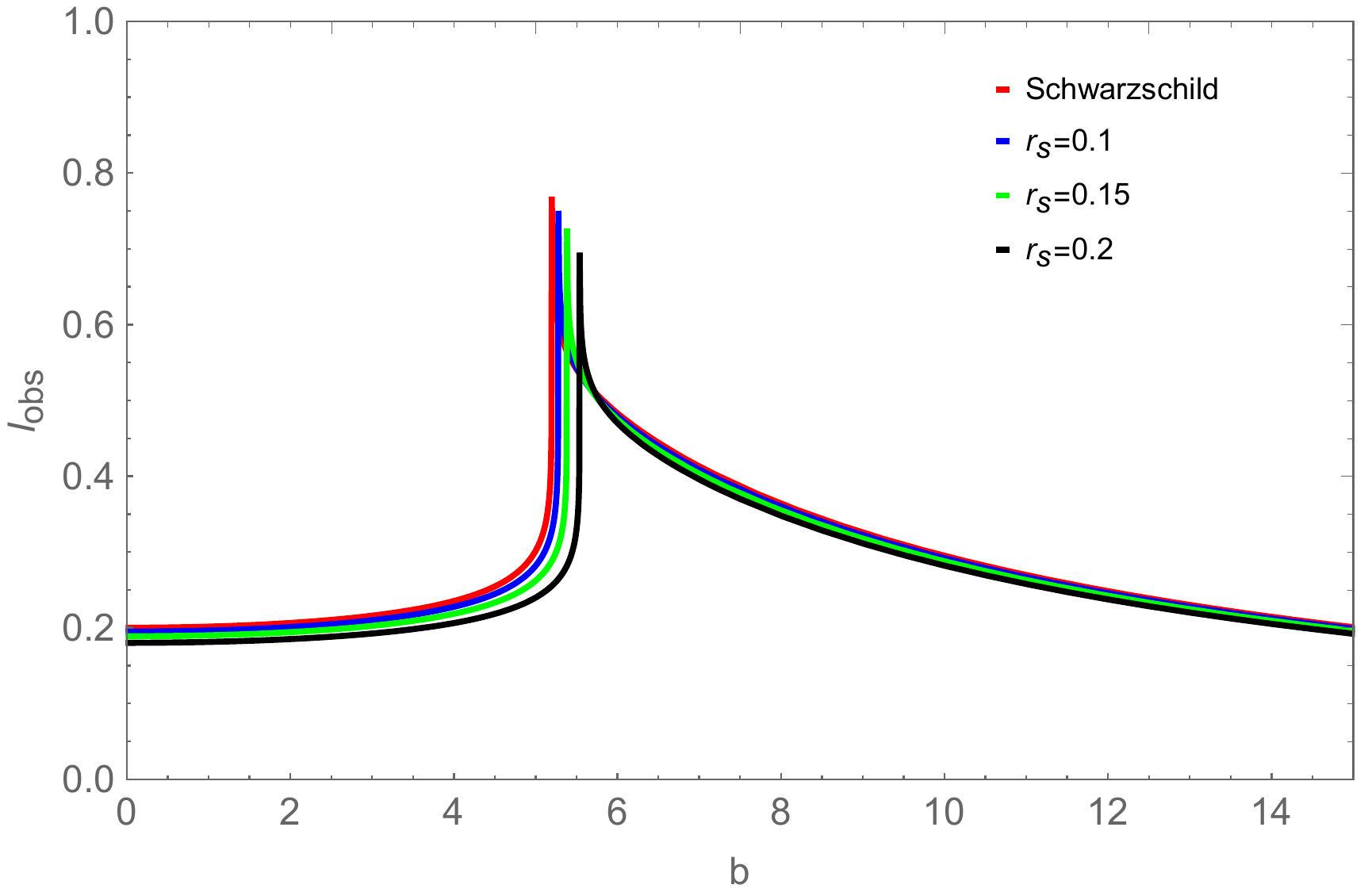}
	\includegraphics[width=0.46\textwidth]{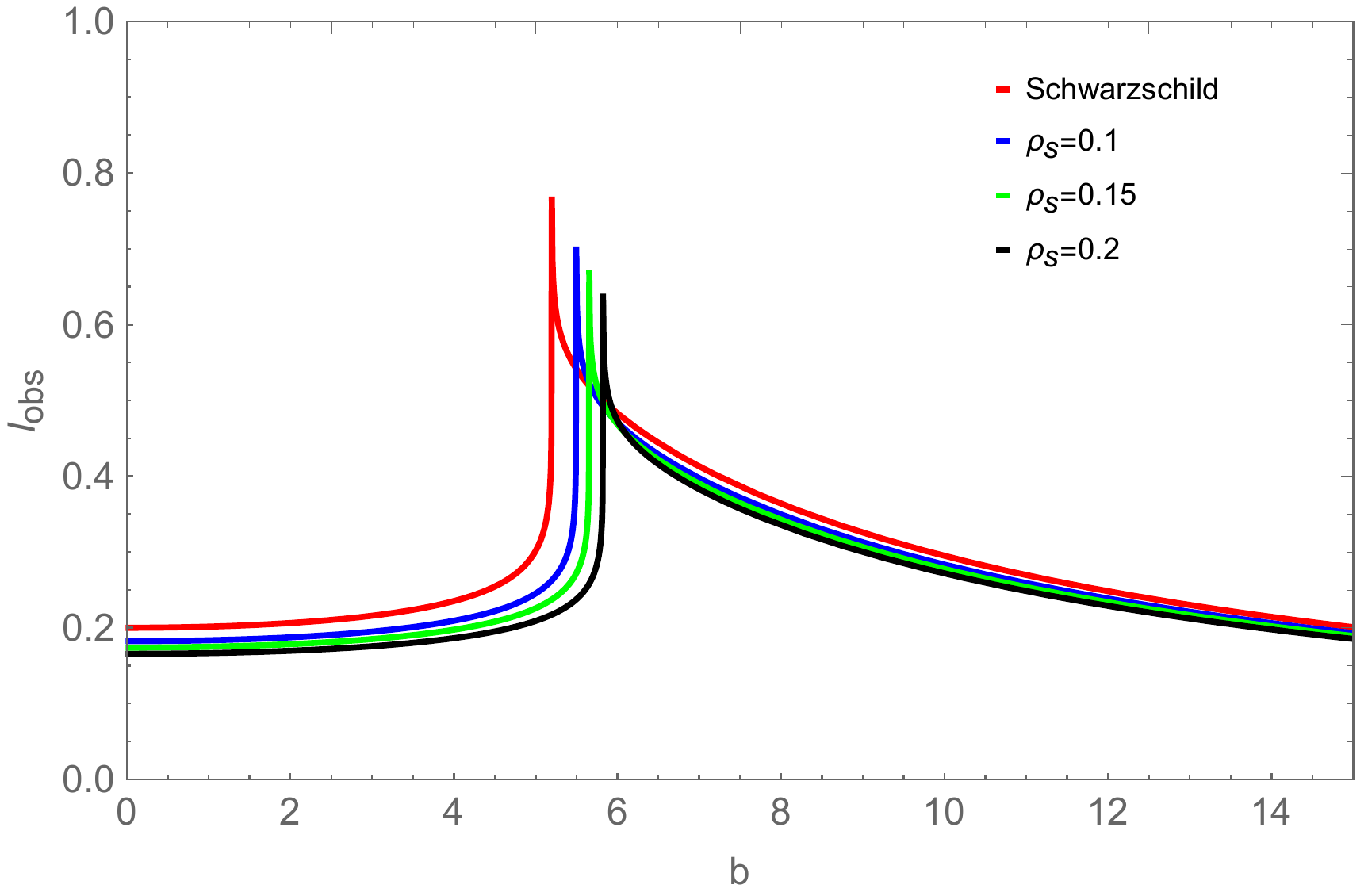}
	\caption{The total observed intensity changes with the impact parameter in the context of the static spherical accretion model. On the left side are the observed intensities when the DM halo density $\rho_s = 0.01$ and $r_s$ takes the values of 0.1 (blue curve), 0.15 (green curve), 0.2 (black curve), and Schwarzschild (red curve) case. On the right side are the observed intensities when $r_s = 0.06$ and $\rho_s $ takes the values of 0.1 (blue curve), 0.15 (green curve), 0.2 (black curv), and Schwarzschild (red curve) case.} 
	\label{fig_011}
\end{figure*}

We first consider a simplified scenario of thin spherical accretion, where the accreting matter is assumed to be statically distributed around this Schwarzschild-like BH.
The radiation intensity observed by an observer at infinity, when a BH is surrounded by thin spherical accretion, can be calculated by integrating the specific intensity of photon radiation along the path 
$ \gamma$ \cite{Bambi:2013nla,Jaroszynski:1997bw}
\begin{equation}\label{31}
	\begin{split}	
		I(\nu_0)=\int_{\gamma} g^{3} j_{e}(\nu_{e}) dl_{prop}
	\end{split},
\end{equation} 
the redshift factor, denoted as $g=\nu_0/\nu_e$, relates the observed frequency \( \nu_0 \) to the emitted frequency \( \nu_e \). Here, \( j_e(\nu_e) \) represents the emissivity per unit volume in the rest frame, which is often approximated as \( j_e(\nu_e) \propto \frac{\delta(\nu_e - \nu_{\star})}{r^2} \), where \( \nu_{\star} \) is the frequency in the rest frame of the source.
$dl_{prop}$ is an infinitesimal proper length. Integrating Eq.~\eqref {31} over all observed frequencies gives the total observed intensity
\begin{equation}\label{iob1}
	\begin{split}	
		I_{obs}=\int_{\nu_0}I(\nu_0)d\nu_0=\int_{\nu_e}\int_{\gamma}g^{4}j_{e}(\nu_{e})dl_{prop}d\nu_{e}
	\end{split}.
\end{equation} 
For a static spherical accretion model, the redshift factor is given by $g = f(r)^{1/2}$, where $f(r)$ is a function of the radial coordinate $r$. Additionally, the infinitesimal proper length $dl_{prop}$ can be expressed as
\begin{equation}\label{33}
	\begin{split}	
		dl_{prop}=\sqrt{\frac{1}{f(r)}dr^{2}+r^{2}d\phi^{2}}=\sqrt{\frac{1}{f(r)}+r^{2}\left(\frac{d\phi}{dr}\right)^{2}}dr
	\end{split},
\end{equation} 
therefore, the cumulative intensity detected by an observer positioned at infinity can be derived as follows:
\begin{equation}\label{34}
	\begin{split}	
		I_{obs}=\int_{\gamma}\frac{f(r)^{2}}{r^{2}}\sqrt{\frac{1}{f(r)}+r^{2}\left(\frac{d\phi}{dr}\right)^{2}}dr
	\end{split},
\end{equation} 
here, $\frac{d\phi}{dr}$ can be obtained from the equation of motion (Eq.~\eqref {rphi}) of the photon in the gravitational field. Then, substitute it into Eq.~\eqref {34}, we have
\begin{equation}\label{35}
	\begin{split}	
		I_{obs}=\int_{\gamma}\frac{g^{4}}{r^{2}}\sqrt{\frac{1}{f(r)}+\frac{b^{2}}{r^{2}-b^{2}f(r)}}dr
	\end{split}.
\end{equation} 
Based on this, it is evident that the total observed intensity is associated with the impact parameter as well as the integration path. Next, in Fig. \ref{fig_011}, we illustrate the variation of the total observed intensity with respect to the impact parameter.

\begin{figure*}[htb]
	\includegraphics[width=0.46\textwidth]{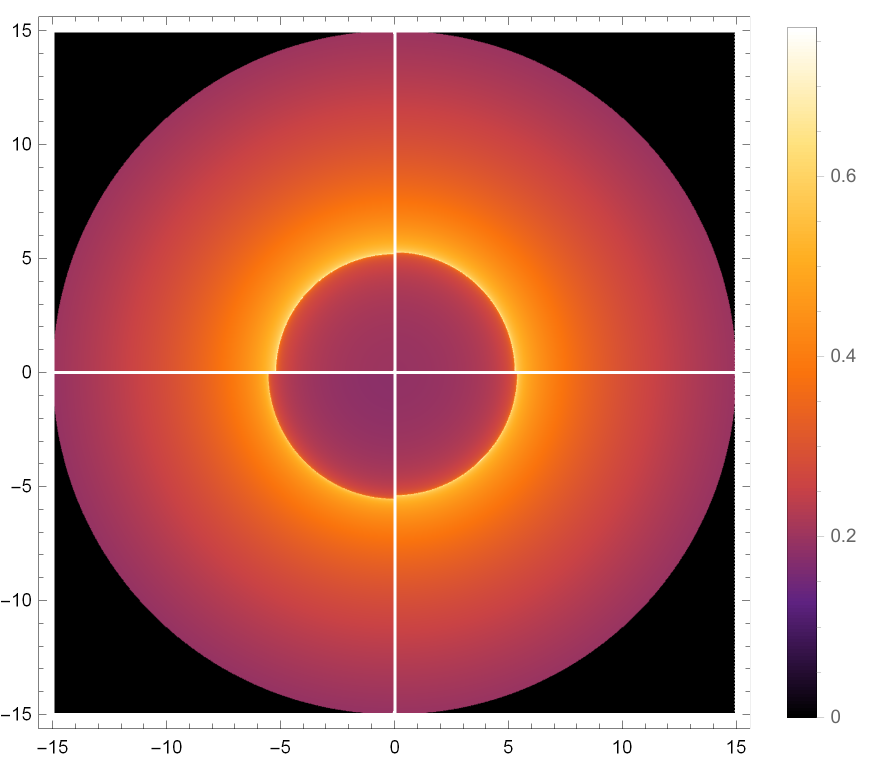}
	\includegraphics[width=0.46\textwidth]{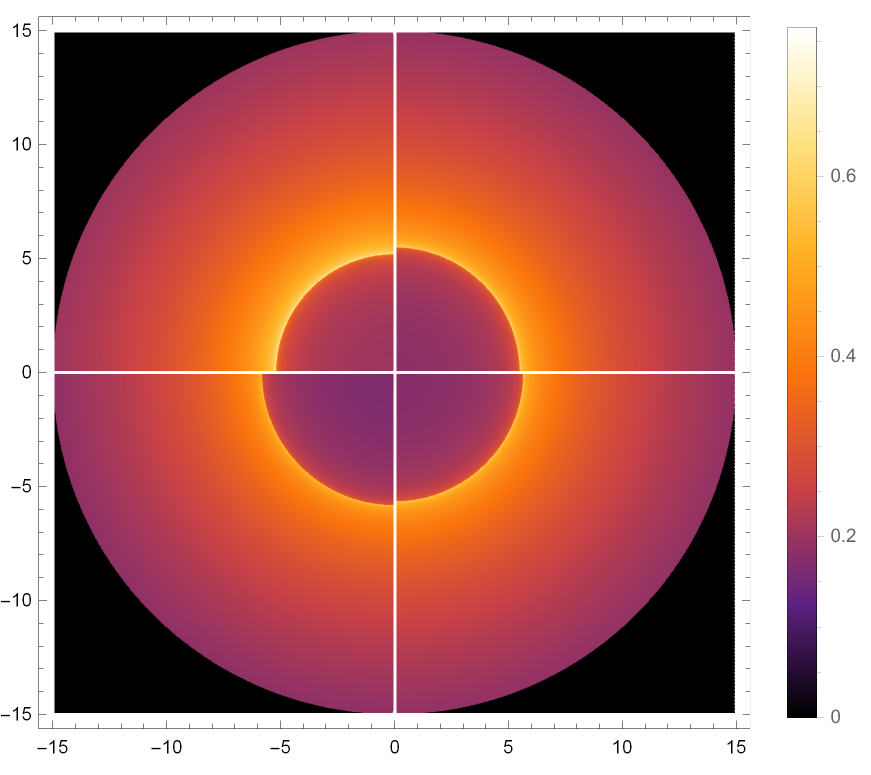}
	\caption{The BH images illuminated by the static spherical accretion.  The left panel compares the optical appearances under the following configurations: (i) Schwarzschild BH (top left); (ii) Schwarzschild-like BHs with fixed $\rho_s=0.01$: $r_s=0.1$ (top right), $r_s=0.15$ (bottom right), and $r_s=0.2$ (bottom left). The right panel contrasts the optical appearances for: (i) Schwarzschild BH (top left); (ii) Schwarzschild-like BHs with fixed $r_s=0.06$: $\rho_s=0.1$ (top right), $\rho_s=0.15$ (bottom right), and $\rho_s=0.2$ (bottom left).  } 
	\label{fig_012}
\end{figure*}

In Fig.~\ref{fig_011}, when one of the DM halo parameters (either $\rho_s = 0.01$ or $r_s = 0.06$) is fixed, the observed intensity steadily diminishes as the other parameter increases, and the overall curve shifts to the right. The peak value of the observed intensity appears at the position where $b=b_{ph}$, and as the variable gradually increases, the position where the peak value appears also shows a trend of gradually increasing. At the peak position, photons orbit multiple times to generate the maximum intensity. As the parameter increases, the observed intensity gradually diminishes, indicating that the brightness of the BH image decreases correspondingly with increasing parameter values.  Furthermore, the BH image illuminated by spherical accretion is presented in Fig. \ref {fig_012}. 

As shown in Fig.~\ref{fig_012}, the faintly luminous central region corresponds to the BH shadow, while the adjacent brighter area represents the spherical accretion region. We observe that as the parameters gradually increase, the shadow region exhibits a clear expansion trend, accompanied by a progressive dimming of the surrounding bright ring. These features demonstrate distinct optical appearances compared to the Schwarzschild BH case, which is fully consistent with our previous analysis.  

\subsection{\label{sec:level5.2}Optical appearance of Schwarzschild-like black hole with radially infalling spherical accretions}

Next, we consider a more general scenario where the Schwarzschild-like BH is surrounded by radially infalling spherical accretion. We adopt the same emission profile as in the static spherical accretion case, which implies that Eq.~\eqref{iob1} remains applicable. For the infalling accretion, the redshift factor can be expressed as \cite{Bambi:2013nla,Wang:2023vcv,KumarWalia:2024yxn}:
\begin{equation}\label{redshift}
	\begin{split}	
		g=\frac{P_{\mu} u_{o}^{\mu}}{P_{\nu} u_{e}^{\nu}}.
	\end{split}
\end{equation}
Here, $P_{\mu}$ denotes the 4-momentum of photons emitted by the accreting matter, while $u_{o}^{\mu}$ and $u_{e}^{\mu}$ represent the 4-velocities of the observer and emitting matter respectively. 
For a distant observer, the 4-velocity is given by $u_{o}^{\mu}=(1,0,0,0)$, while for radially infalling accreting matter, the 4-velocity can be expressed as
\begin{equation}\label{4su}
	u_e^t=\frac{1}{f(r)}, \qquad u_e^r=-\sqrt{1-f(r)}, \qquad u_e^\theta=u_e^\phi=0.
\end{equation}
The photon 4-momentum $P_{\mu}$ can be derived from the Lagrangian via $P_{\mu} = \frac{\partial \mathcal{L}}{\partial \dot{x}^{\mu}}$, yielding
\begin{equation}\label{monmentum}
	\begin{split}	
		P_t&=-\frac{1}{b}, \\  P_r&=\pm \frac{1}{f(r)}\sqrt{\frac{1}{b^2}-\frac{f(r)}{r^2}},
	\end{split}
\end{equation}
where the $\pm$ sign indicates the photon moving inward or outward relative to the Schwarzschild-like BH.
Note that here we only display the two required components of the photon 4-momentum. By substituting Eqs.~\eqref{4su} and~\eqref{monmentum} into Eq.~\eqref{redshift}, we can therefore calculate the corresponding redshift factor 
\begin{equation}\label{redshift1}
	g=\frac{1}{\frac{1}{f(r)}\pm \sqrt{\frac{1}{f(r)}\left[\frac{1}{f(r)}-\frac{b^2}{r^2}\right] [1-f(r)]}}.
\end{equation}
Regarding the proper length $dl_{prop}$, it is defined as \cite{Bambi:2013nla,Wang:2023vcv}
\begin{equation}\label{guyou}
	dl_{prop}=P_\mu u_e^\mu d\lambda=\frac{P_t}{g|P_r|}dr.
\end{equation}
Finally, by combining Eqs.~\eqref{iob1} and~\eqref{guyou}, the total observed intensity can be calculated through the following expression \cite{Bambi:2013nla,Wang:2023vcv}
\begin{equation}\label{iob3}
	I_{obs}=\int_{\gamma}  \frac{g^3 P_t}{r^2 |P_r|} dr.
\end{equation}

\begin{figure*}[htb]
	\includegraphics[width=0.46\textwidth]{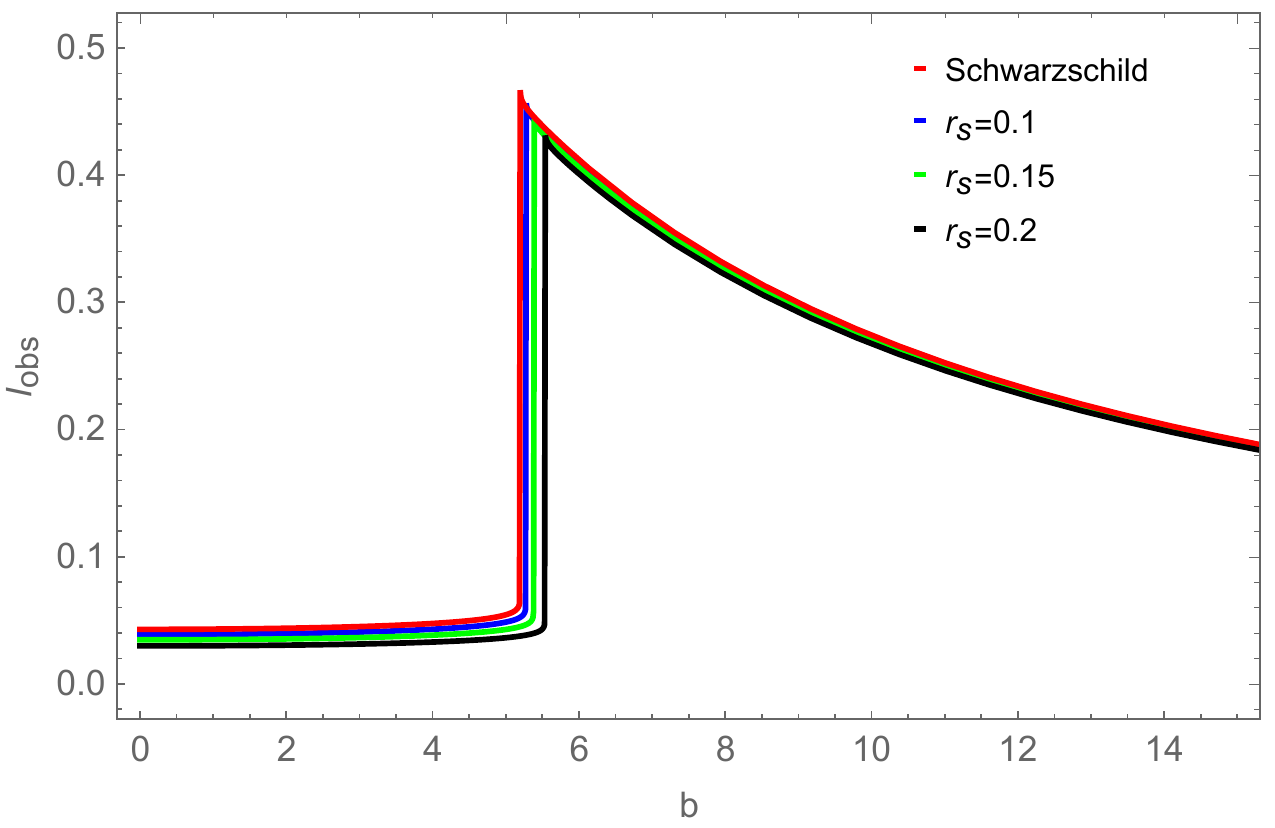}
	\includegraphics[width=0.46\textwidth]{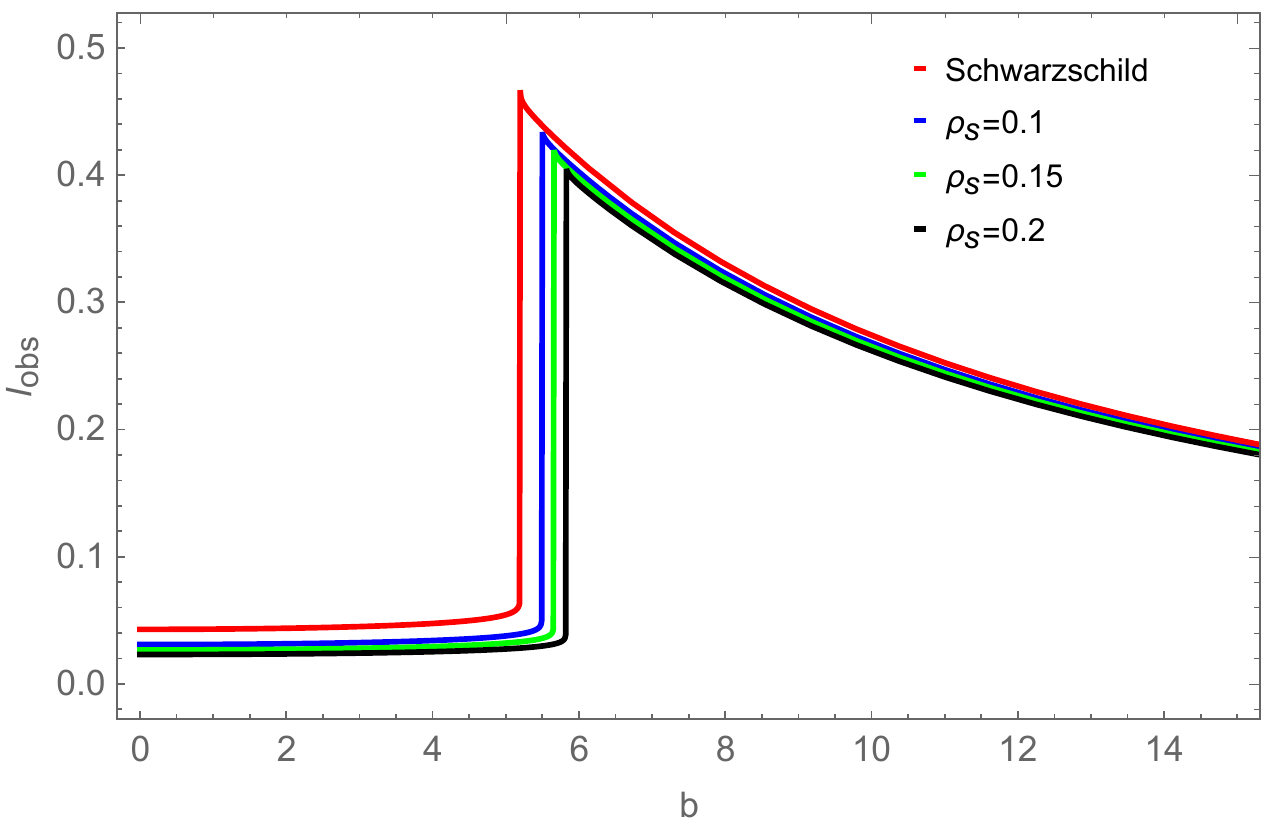}
	\caption{The variation of total observed intensity with impact parameter in the radially infalling spherical accretion. The left panel displays the received intensity for cases with fixed DM halo density $\rho_s = 0.01$, where $r_s$ takes values of 0.1, 0.15, 0.2 and the Schwarzschild case, represented by blue, green, black and red curves respectively. The right panel shows the corresponding intensity comparison for fixed $r_s = 0.06$ with $\rho_s$ values of 0.1, 0.15, 0.2 and the Schwarzschild case, similarly depicted using the same color scheme.}
	\label{infalling_iob}
\end{figure*}

In Fig.~\ref{infalling_iob}, we present the received intensities for both Schwarzschild and Schwarzschild-like BHs under different DM parameter configurations. The left panel of Fig.~\ref{infalling_iob} displays the results with fixed $\rho_s=0.01$ for $r_s=0.1$, $0.15$, and $0.2$, while the right panel shows the cases with fixed $r_s=0.06$ for $\rho_s=0.1$, $0.15$, and $0.2$. Notably, the parameter-dependent variations in received intensity follow similar trends to those observed in static spherical accretion. 

\begin{figure*}[htb]
	\includegraphics[width=0.46\textwidth]{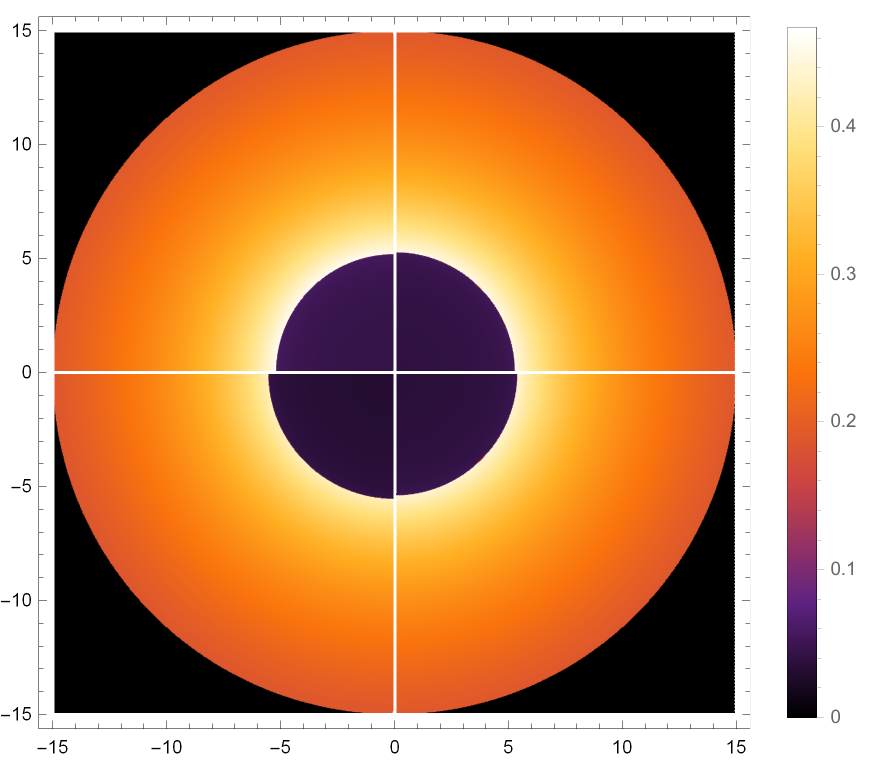}
	\includegraphics[width=0.46\textwidth]{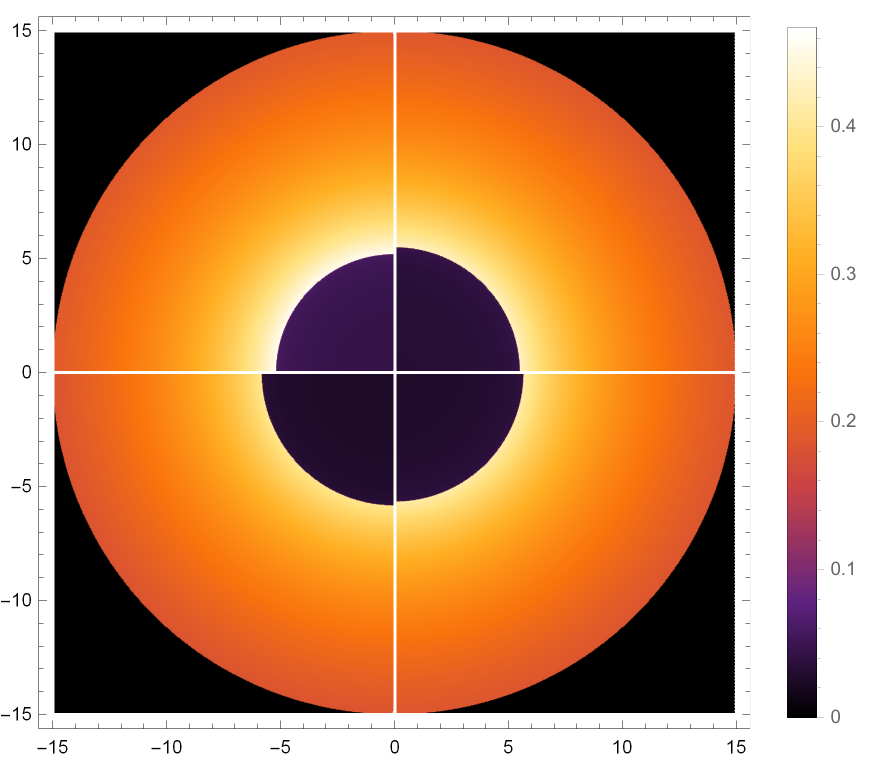}
	\caption{The BH images illuminated by the radially infalling spherical accretion. The left panel compares the optical appearances under the following configurations: (i) Schwarzschild BH (top left); (ii) Schwarzschild-like BHs with fixed $\rho_s=0.01$: $r_s=0.1$ (top right), $r_s=0.15$ (bottom right), and $r_s=0.2$ (bottom left). The right panel contrasts the optical appearances for: (i) Schwarzschild BH (top left); (ii) Schwarzschild-like BHs with fixed $r_s=0.06$: $\rho_s=0.1$ (top right), $\rho_s=0.15$ (bottom right), and $\rho_s=0.2$ (bottom left).  } 
	\label{infalling_image}
\end{figure*}

The corresponding optical appearances of BH surrounded by the infalling spherical accretion are shown in Fig.~\ref{infalling_image}. Our analysis reveals that the shadow radius observed in the viewer's frame consistently corresponds to the critical impact parameter for both static and radially infalling accretion scenarios. Moreover, the optical appearance of the infalling spherical accretion exhibits identical dependence on DM parameters as its static counterpart--both demonstrate progressively larger and darker central shadow regions with increasing DM parameter values. These results demonstrate that the optical appearance of BHs illuminated by different accretion models provides an effective observational diagnostic to distinguish Schwarzschild-like BHs from their classical Schwarzschild counterparts.


\section{\label{sec:level6}Conclusion}
In this paper, we have studied the images of Schwarzschild-like BH surrounded by Dehnen-type DM halo under different accretion models, mainly analyzing the impacts of the density and radius of DM halos on the optical appearance. We first discussed the photon geodesics and the effective potential in the spacetime of the Schwarzschild-like BH. Our calculations revealed that the peak value of the effective potential decreases monotonically as $\rho_s$ and $r_s$ increase, while its radial coordinate increases accordingly. Then, we constrained the DM halo parameters $\rho_{s}$ and $r_{s}$ using the EHT measured shadow radius of M87*. 

Subsequently, within the constraint range of M87*, we systematically investigated photon trajectories across varying $\rho_{s}$ and $r_{s}$ values. The results demonstrate that an increase in $\rho_{s}$ and $r_{s}$ leads to an increase in the impact parameters related to the lensing ring and photon ring. Furthermore, our analysis revealed that the Schwarzschild-like BH's the critical impact parameter $b_{ph}$, the radius $r_{isco}$ of the innermost stable circular orbit, the event horizon radius $r_h$, and the radius $r_{ph}$ of the photon sphere all exhibit monotonically increasing trends as these parameters increase. Moreover, we investigated the variations of the first three transfer functions with DM halo parameters, finding that these transfer functions systematically shift toward larger values of the impact parameter $b$ with increasing $\rho_{s}$ and $r_{s}$.

Then, we investigated the optical appearances of a Schwarzschild-like BH illuminated by optically thin accretion disks. Through three simplified models, we found that different models exhibit significant differences in the distribution of observed intensity and the resulting optical appearance, as shown in Figs.~\ref{fig:rs} and ~\ref{fig:ros}. The observed intensity profile can be decomposed into three distinct components: direct emission, lensing ring, and photon ring contributions. Moreover, the DM halo parameters $\rho_{s}$ and $r_{s}$ significantly affect the distribution of bright rings in the optical image. As $\rho_{s}$ and $r_{s}$ increase, the associated characteristic radius also increase, leading to an expansion of the image radius and causing the bright rings to appear farther from the BH, thereby producing optical features that deviate from those of a Schwarzschild BH.

Finally, we further examined the optical characteristics of a Schwarzschild-like BH under both static and radially infalling spherical accretion scenarios (see Figs.~\ref{fig_012} and ~\ref{infalling_image}). The findings indicate that, akin to the case of an optically thin disk, the peak of the observed intensity moves to higher values of the impact parameter $b$ as $\rho_{s}$ and $r_{s}$ grow. However, the observed intensity gradually decreases, resulting in a diminished overall brightness of the optical image. These findings provide a valuable reference for understanding the variation patterns of the optical features of Schwarzschild-like BHs under the combined influence of specific accretion environments and DM halo parameters. 

It is worth noting that extensive research has been conducted on BH solutions surrounded by various DM models \cite{Al-Badawi:2025njy,Konoplya:2022hbl,Konoplya:2025mvj,Macedo:2024qky,Li:2024abk,Haroon:2025rzx,Pantig:2022whj,Liu:2023oab,Liu:2022ygf}. In studies concerning the shadows and optical appearance of BHs surrounded by DM \cite{Li:2024abk,Zeng:2025kqw,He:2024amh,Jha:2025xjf,Jha:2024ltc,Al-Badawi:2024qpt,Hou:2018avu,Xu:2018mkl}, the Schwarzschild BH embedded in Dehnen-(1, 4, 5/2) type DM—as well as its counterpart in the presence of a quintessence field \cite{Hamil:2025pte}—exhibits a gradual increase in both shadow size and optical appearance as the DM parameter grows. This behavior aligns with shadow variations induced by certain DM models, such as cold DM \cite{Zeng:2021mok,Zeng:2025kqw}. In contrast, some DM models—for instance, Hernquist-type DM \cite{Macedo:2024qky}—demonstrate a decrease in shadow size with increasing DM parameter, exhibiting trends distinct from those of the Dehnen-type model. These observations suggest that the optical appearance of a Schwarzschild-like BH illuminated under different accretion disk models may allow us to distinguish between various DM profiles based on image characteristics. For models exhibiting similar shadow behavior under parameter variation, further in-depth analysis will be required. A systematic comparison of optical appearances across different DM scenarios may be pursued in future research.

In conclusion, our study demonstrates the influence of the Schwarzschild-like BH parameters, when surrounded by a Dehnen-type DM halo, on its optical appearance under different accretion models. Our results suggest that the optical appearance of a Schwarzschild-like BH exhibits significant differences compared to its Schwarzschild counterpart. This offers a preliminary perspective on the properties of BHs surrounded by DM and suggests a possible method for distinguishing these BHs in future observations.

\section{Acknowledgements}

This work was supported by Guizhou Provincial Basic Research Program (Natural Science) (Grant No. QianKeHeJiChu-[2024]Young166),  the National Natural Science Foundation of China (Grant No. 12365008) and the Guizhou Provincial Basic Research Program (Natural Science) (Grant No.QianKeHeJiChu-ZK[2024]YiBan027 and QianKeHeJiChu-MS[2025]680.).

\bibliography{ref}

\end{document}